\documentclass[
10pt,
amsmath,amssymb,
aps,
prb,
longbibliography,
onecolumn,
superscriptaddress
]{revtex4-2}
\usepackage{soul}
\usepackage[utf8]{inputenc}
\usepackage[T1]{fontenc}
\usepackage{orcidlink}
\usepackage{amssymb}
\usepackage{amsmath}
\usepackage{graphicx}
\usepackage{bm}
\usepackage{xcolor}
\usepackage{enumitem}
\usepackage{array}
\usepackage{hyperref}
\usepackage{listings}
\usepackage{comment}
\usepackage{setspace}
\usepackage{datetime}
\usepackage{booktabs}
\usepackage{mathtools}
\usepackage{slashed}

\newcommand{\be}{\begin{equation}}
\newcommand{\ee}{\end{equation}}
\newcommand{\ham}{\mathsf{H}}
\newcommand{\hamop}{\widehat{\mathsf{H}}}
\newcommand\operator[1]{\widehat{\mathsf{#1}}}

\newcommand{\vc}{\mathsf{c}}
\newcommand{\vcop}{\widehat{\mathsf{c}}}
\newcommand{\vaop}{\widehat{\mathsf{a}}}
\newcommand{\vAop}{\widehat{\mathsf{A}}}
\newcommand{\vbop}{\widehat{\mathsf{b}}}
\newcommand{\vBop}{\widehat{\mathsf{B}}}
\newcommand{\vR}{\mathbf{R}}
\newcommand{\vn}{\widehat{\mathsf{n}}}
\newcommand{\vd}{\mathsf{d}}
\newcommand{\vj}{\mathsf{j}}
\newcommand{\la}{\langle}
\newcommand{\ra}{\rangle}
\newcommand{\p}{\partial}

\begin{document}

\title{Instantons and topological order in two-leg electron ladders: A universality class}

\author{S.-R. Eric Yang\,\orcidlink{0000-0003-3377-1859}}
\thanks{Corresponding author: eyang812@gmail.com}
\affiliation{%
Department of Physics, Korea University, Seoul 02841, Republic of Korea
}%

\author{Hyun Cheol Lee\,\orcidlink{0000-0003-1352-8851}}
\affiliation{%
Department of Physics, Sogang University, Seoul 04107, Republic of Korea
}%

\author{Hoang-Anh Le\,\orcidlink{0000-0002-1668-8984}}
\affiliation{%
Center for Quantum Nanoscience, Institute for Basic Science, Seoul 03760, Republic of Korea
}%
\affiliation{%
Ewha Womans University, Seoul 03760, Republic of Korea
}%

\author{In-Hwan Lee\,\orcidlink{0000-0002-7619-3780}}
\affiliation{%
Department of Physics, Korea University, Seoul 02841, Republic of Korea
}%

\begin{abstract}
Our numerical study of the disordered Hubbard model with nearest-neighbor hopping shows that a two-leg electron ladder has a finite topological entanglement entropy in the regime where the density of states exhibits an exponentially decaying gap. The value of the topological entanglement entropy suggests that two-leg ladders belong to the same universality class as graphene zigzag nanoribbons, despite several structural differences. A Shankar-Witten-type bosonization Lagrangian with disorder captures several features of the numerically obtained results for disordered two-leg ladders. Additionally, we propose a Lagrangian in which the fusion of two semions residing on different chains generates a fermion (instanton).  We apply this Lagrangian within the framework of the pinned charge density wave model and compute the relevant Green's function using the bosonization method.
This approach predicts a linear density of states at a critical disorder strength. Below this threshold, a soft gap emerges, which is in qualitative agreement with our numerical results.

\noindent{\it Keywords:  Two-Leg Ladder, Topological Order,  Instanton, Semion, Fractional Charge}
\end{abstract}

\maketitle

\section{Introduction}

Well-known topologically ordered insulators include spin liquids, integer, fractional, and anomalous quantum Hall states, superconductors, and graphene zigzag nanoribbons (for reviews, see \cite{wen2017, HaldaneNobel, GV2019, yangbook}).  Some of them support Abelian anyons, while others exhibit non-Abelian anyons or none at all (with anyons obeying neither Fermi nor Bose statistics). Nevertheless, they all exhibit long-range entanglement, and some possess a nonzero topological entanglement entropy~\cite{Kitaev2006,Wen2006}, revealing an intriguing  connection between the edge and entanglement spectra~\cite{Haldane2008}.
The stability of anyons with fractional charges appears to be intimately related to the well-defined nature of topological entanglement entropy~\cite{Le_2024}.
In this paper, we propose that disordered two-leg ladder lattices can also be topologically ordered insulators with charged semions. Semions~\cite{Haldane1991, Kalmeyer1987} have recently attracted significant interest as they may contribute to the development of superconductivity~\cite{Laughlin1988, Canright1989, Zee1990}.
		
In the absence of disorder, interacting two-leg spin ladders can exhibit rich physics; for reviews, see Refs.~\cite{Haddad2000, GRUNINGER20022167, Johnston2000}.
In a two-leg spin-1/2 ladder, rung singlets can naturally form. If these singlets interact collectively, they can generate effective spin-1 excitations at lower energy scales.
 This can lead to phenomena similar to the Haldane-like gap in spin-1 chains.
 However, a two-leg ladder with open boundaries does not necessarily host fractionalized end states as in the Haldane chain~\cite{Haldane1983}, which is an example of symmetry-protected topological insulator. 
If the rungs are weakly coupled, the system behaves more like two weakly coupled gapless spin-Heisenberg chains, and the Haldane gap does not appear.
A two-leg spin-1/2 ladder is, therefore not a symmetry-protected topological insulator.

Can a quasi-one-dimensional two-leg electron ladder described by a Hubbard Hamiltonian, rather than a spin Hamiltonian, exhibit a topological phase? Recently, it was shown that, in an incommensurate potential, a two-leg electron ladder with inversion symmetry may exhibit topological phases~\cite{Agh_2024}. However, the disorder breaks inversion, mirror, rotational, and chiral symmetries. The presence of magnetism also breaks time-reversal symmetry.
Moreover, the addition of an impurity has a singular effect on interacting one-dimensional systems, effectively dividing the system into two domains~\cite{Kane1992, Furusake1993}. This is analogous to how a soliton connects two regions with different dimerized phases in polyacetylene, as described in~\cite{HeegerMod1988}.
Because of interchain hopping, these solitons may form instantons, similar to those in coupled quantum double wells, where the right (left) well corresponds to the upper (lower) chain of the ladder. Instantons are known to play a crucial role in generating topological order in zigzag graphene nanoribbons~\cite{yangbook}.
Another reason why gapped two-leg electron ladder systems are intriguing in the context of topologically ordered insulators is that, unlike ordinary topological materials, their boundary edges and bulk are not distinctly separable. The upper (lower) boundary consists of the upper (lower) leg. 
It is unclear how these effects may influence topological properties.

We introduce disorder and numerically study a disordered Hubbard model on a two-leg electron ladder with nearest-neighbor hopping. The following are the main results of our study.
We show that disorder fragments the ladder into distinct regions, with boundaries separating neighboring segments. Instantons—each consisting of a pair of spatially separated fractional charges—emerge at these boundary locations, linking corresponding points on the opposite chains, as originally proposed in Ref.~\cite{jeong2019}.
 These fractional charges reside on the side boundaries rather than on the end sites of the ladder. We argue that an effective Lagrangian of the Shankar-Witten~\cite{Shankar1978} type, incorporating disorder, can explain the obtained numerical results and provide further insight.
We also demonstrate that disordered, interacting two-leg ladders exhibit a nonzero topological entanglement entropy in the regime where the density of states exhibits an exponentially decaying gap.  
Using the bosonization approach, we propose a Lagrangian incorporating semion phase fields for the two-leg ladder. In this model, an instanton (fermion) is mathematically well defined when two semions on opposite chains fuse. The electron Green's function, calculated from this Lagrangian using the charge-density wave (CDW) pinning model, suggests either a linear density of states (DOS) or a soft gap, depending on the strength of the disorder, in qualitative agreement with the numerical results. This supports the presence of semions in topologically ordered two-leg ladders.
The physical properties of two-leg ladders and zigzag graphene nanoribbons differ in several respects. However, they exhibit the same fractional charge values and an approximately identical universal value of topological entanglement entropy within numerical uncertainty. We believe that two-leg ladders and zigzag graphene nanoribbons belong to the same universality class~\cite{Pachos_2012}.

This paper is organized as follows. Before examining the disordered and interacting two-leg ladder, we first numerically study disordered one-dimensional quantum chains in Sec.~\ref{section:OneChain}. These systems fragment into regions with distinct magnetic orders, connected by soliton (kink) states.
In Sec.~\ref{section:TwoLeg}, we numerically investigate disordered, interacting two-leg ladder lattices. When an exponentially decaying gap develops, the topological entanglement entropy becomes well defined. These systems also exhibit instantons carrying $e/2$ fractional charges.
To better understand the role of magnetic order, kinks, and instantons in the numerical results, we discuss the Shankar-Witten-type Lagrangians in Sec.~\ref{section:Bosonization1D}.  In Sec.~\ref{section:Bosonization2Leg}, we examine how the fusion of semions gives rise to an electron in the topologically ordered two-leg ladder, following the bosonization framework for quasiparticle operators in quantum Hall edge states.
In Sec.~\ref{sec:cdw}, we compute the density of states (DOS) within the charge-density wave (CDW) pinning model using the electron Green's function derived from the fusion of semions. This method is then used to compute both a linear gap and a soft gap in the DOS, which qualitatively agree with numerical results.
The similarities and differences between zigzag graphene nanoribbons and two-leg ladders are analyzed in Sec.~\ref{section:simanddiff}, indicating a universality class. Finally, conclusions and discussions are presented in Sec.~\ref{section:discussion}.


\section{Numerical results of disordered one-dimensional quantum chain}
\label{section:OneChain}

Before we study two-leg ladder systems, let us first investigate a simple self-consistent mean-field Hamiltonian of a disordered one-dimensional chain at {\it half-filling}. 
We include terms consisting of on-site repulsion $U$ and on-site disorder potential $V_{l}$ 
\begin{eqnarray}
\hamop=&-t\sum_{l,\sigma} \left[ \vcop_{l,\sigma}^{\dag} \vcop_{l+1,\sigma}+\text{H.c.} \right]  +\sum_{l,\sigma} V_{l} \, \vcop_{l,\sigma}^{\dag} \vcop_{l,\sigma} \nonumber\\
&+U\sum_{l}\left[ \vn_{l,\uparrow}\langle \vn_{l,\downarrow}\rangle +\vn_{l,\downarrow}\langle \vn_{l,\uparrow}\rangle-\langle \vn_{l,\downarrow}\rangle\langle \vn_{l,\uparrow}\rangle \right].
\label{PolyHam0}
\end{eqnarray}
Here, $\vcop_{l,\sigma}$ is the electron destruction operator for site $l$ and spin $\sigma=\uparrow$ or $\downarrow$,
and $\vn_{l,\sigma}$ is the occupation number operator. $t$ represents the hopping amplitude. 
We introduce diagonal disorder $V_l$ with a concentration of $n_{\rm imp} = 0.1$, where $V_l \in [-\Gamma, \Gamma]$ follows a Gaussian distribution. 
Note that the spin-up and spin-down components of the above Hamiltonian separate:   $\hamop=\hamop_{\uparrow}+\hamop_{\downarrow}$.

The Hartree-Fock approach is often reliable in systems with an energy gap. Additionally, Anderson localization tends to suppress quantum fluctuations. In our previous study (Ref.~\cite{yang2022}) on the narrowest zigzag graphene nanoribbons, conducted using the DMRG method, we found that the Hartree-Fock results showed qualitative agreement with the DMRG findings. Additionally, our Hartree-Fock results for the one-dimensional chain are consistent with those of Refs.~\cite{Kane1992, Furusake1993}, which are based on the bosonization.

In the absence of disorder, the length of the \textit{magnetic unit cell} (the repeating structure of the magnetic order) is twice that of the lattice period. 
It is useful to define the local antiferromagnetic  order parameter as follows 
(below $\operator{S}_z=(\vn_{l\uparrow}-\vn_{l\downarrow})/2$ is the $z$ component of the spin operator): 
\begin{eqnarray}
	\Phi(l) =\frac{1}{M_s} \left\{ \sum_{k  \in \textrm{odd sites}} \Bigl \langle     \operator{S}_z(k) \Bigr \rangle    - \sum_{p  \in \textrm{even sites}} \Bigl \langle \operator{S}_z(p) \Bigr \rangle \right\},
	\label{mag_order_param}
\end{eqnarray}
where $k$ and $p$ denote sites around site $l$, and $M_s$ is the number of these even-numbered or odd-numbered sites.

\begin{figure}[hbt!]
\begin{center}
\includegraphics[width = 0.7\textwidth]{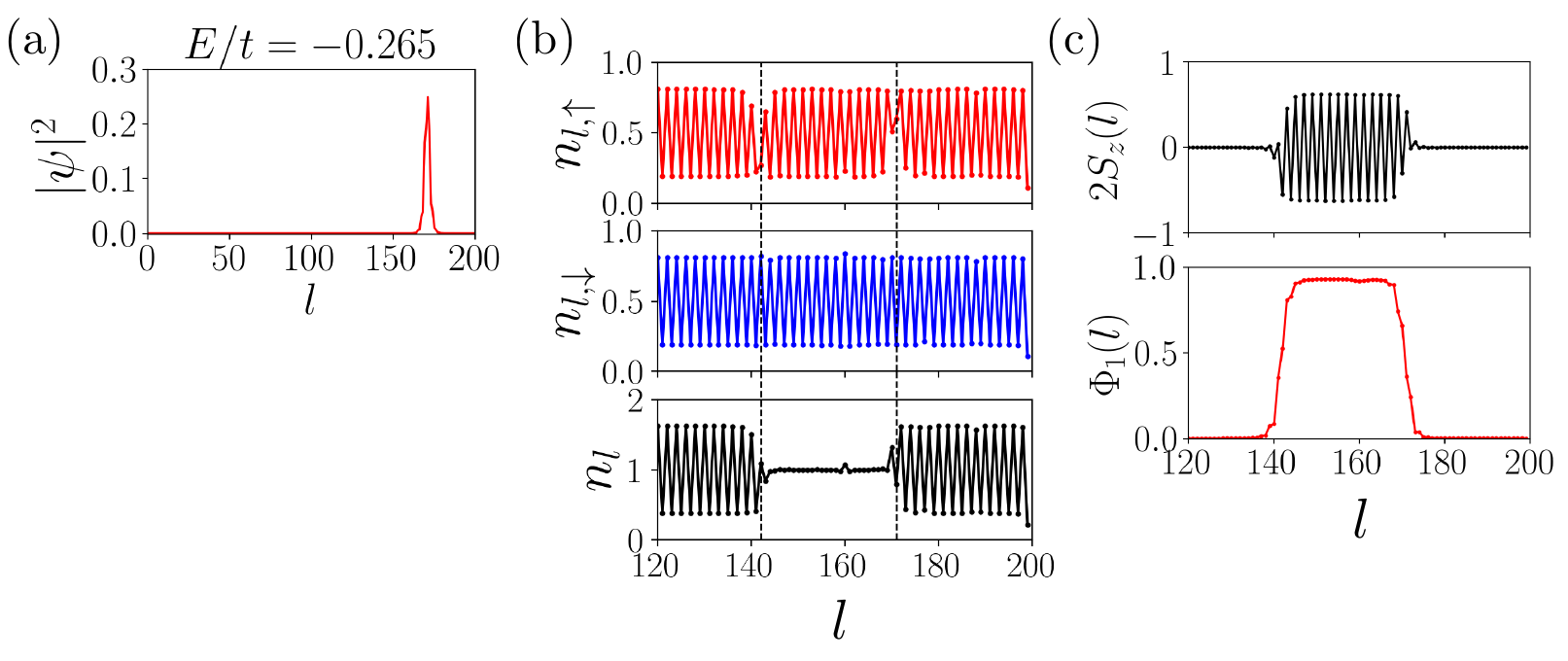}
\caption{(a) The probability density of a state that leads to  the formation of a magnetic kink. (b) Site occupation numbers 
$\langle \vn_{l,\uparrow}\rangle$, 
$\langle \vn_{l,\downarrow}\rangle$, and  
$\langle \vn_{l}\rangle=\langle \vn_{l,\uparrow}\rangle+\langle \vn_{l,\downarrow} \rangle$ are displayed. Vertical lines indicate the positions where the spin-up and/or spin-down occupation numbers change.
(c) Site spin values $2\langle \operator{S}_z\rangle$  are shown. Local magnetic order parameter $\Phi_1(l)$  is also plotted. The parameters $(U,\Gamma,n_\text{imp})=(3t,0.5t,0.1)$ are used with the length $200a_0$, where  $n_{\rm imp}$ is the impurity concentration and  $a_0$ is the lattice period.}
\label{quantum_chain_ground}
\end{center}	
\end{figure}

A kink state is an eigenstate of the Hamiltonian Eq. (\ref{PolyHam0}). Their presence is closely tied to the positions of the onset of the antiferromagnetic domain. A localized gap state is responsible for this effect. The probability density of a kink state is shown in Fig. \ref{quantum_chain_ground}(a).   
The occupation numbers are shown in Fig. \ref{quantum_chain_ground}(b).   Both $\langle \vn_{l,\uparrow}\rangle$ and $\langle \vn_{l,\downarrow}\rangle$ oscillate from site to site. The corresponding site spins $\langle \operator{S}_z\rangle$ are shown in Fig. \ref{quantum_chain_ground}(c). They exhibit a finite-length antiferromagnetic domain.
Note that occupation numbers abruptly change by $1/2$ approximately at several locations where kinks are present.
Fig. \ref{quantum_chain_ground}(c) shows the antiferromagnetic order parameter, with kinks (solitons) occurring where the magnetic order parameter changes abruptly.
Thus, in the presence of disorder, a one-dimensional system is divided into two types of regions: one with magnetic order and the other without. This is a singular effect of disorder~\cite{Kane1992,Furusake1993}. We will see that this also occurs in two-leg ladders.

\section{Numerical results of disordered two-leg ladder  model}
\label{section:TwoLeg}

\subsection{Hubbard model of  two-leg ladder lattices and band structure}

\begin{figure}[hbt!]
\begin{center}
\includegraphics[width = 0.3 \textwidth]{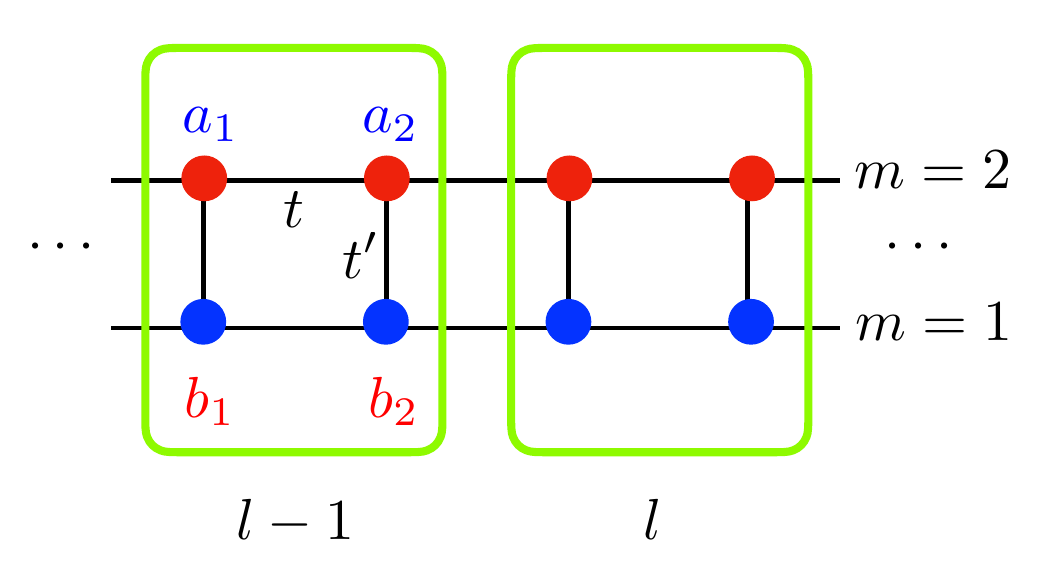}
\caption{A two-leg ladder is shown with  horizontal (intrachain) and vertical (interchain) hopping parameters $t$ and $t'$. The red (blue) chain is denoted by $m=2\,\, (m=1)$.  The green box indicates  a magnetic unit cell.   The magnetic unit-cell length is $a=2a_0$, where $a_0$ is the lattice constant of the  two-leg ladder, i.e., the distance between neighboring atoms on a chain (in momentum space, the unit cell is half as long).}
\label{Schematic_TLL}
\end{center}	
\end{figure}

In Sec.~\ref{section:OneChain}, we considered the effect of disorder in a one-chain model. 
Now, we study a half-filled two-leg ladder, consisting of two coupled chains in the presence of electron interactions. The lattice structure of a two-leg ladder is shown in Fig.~\ref{Schematic_TLL}.  First, we investigate the case in the absence of disorder.

The self-consistent mean-field Hamiltonian can be written as 
\begin{equation}
\begin{aligned}
\hamop_\text{MF} = \; &U \sum_l \sum_p \left[ \vn_{mp\uparrow}(l) \left< \vn_{mp\downarrow}(l) \right> +\vn_{mp\downarrow}(l) \left< \vn_{mp\uparrow} (l)\right>  \right] - t \sum_{l,\sigma} \left[ \vaop^\dagger_{l1\sigma} \vaop_{l-1,2\sigma} + \vaop^\dagger_{l 2 \sigma} \vaop_{l 1 \sigma} + \text{H.c.} \right] \\
&- t \sum_{l,\sigma} \left[ \vbop^\dagger_{l1\sigma} \vbop_{l-1,2\sigma} + \vbop^\dagger_{l 2 \sigma} \vbop_{l 1 \sigma} + \text{H.c.} \right] - t' \sum_{l,\sigma} \left[ \vaop^\dagger_{l1\sigma} \vbop_{l1\sigma} + \vaop^\dagger_{l 2 \sigma} \vbop_{l 2 \sigma} + \text{H.c.} \right].
\end{aligned}
\label{twoleg}
\end{equation}
The index $l$ labels magnetic unit cells (unlike in Eq.~(\ref{PolyHam0})), while $t$ and $t'$ are intra-chain hopping and inter-chain hopping, respectively. The electron destruction operators for the sites on the upper chain ($m=2$) and lower chain ($m=1$) are denoted by $\vaop_{p\sigma}(l)$ and $\vbop_{p\sigma}(l)$, respectively,  where $p=1,2$  labels the left or right atoms in a 
magnetic unit cell.

Let us introduce the momentum $k$-space  destruction operators $\vAop_{p\sigma}(k)$ and $\vBop_{p\sigma}(k)$  via the real space destruction operators as follows: 
\begin{eqnarray}
	\vaop_{lp\sigma}=\frac{1}{\sqrt{L}}\sum_k e^{ik x_{lp}}\vAop_{p\sigma}(k),\nonumber\\
	\vbop_{lp\sigma}=\frac{1}{\sqrt{L}}\sum_k e^{ik x_{lp}}\vBop_{p\sigma}(k),
\end{eqnarray}
where $x_{lp}$  denotes the position of an atom along a chain in the $l$th magnetic unit cell, and $L$ is the length of the chain.     
The  Hartree-Fock mean-field Hamiltonian Eq.~(\ref{twoleg})  separates into each spin component  $\hamop_{\uparrow}$ and $\hamop_{\downarrow}$,
\begin{eqnarray}
	\hamop_{\rm MF}=\hamop_{\uparrow}+\hamop_{\downarrow}.
\end{eqnarray}
The Fourier-transformed Hamiltonian for the electron with spin $\sigma$ is given by
\begin{align}
	\hamop_{\sigma}=\sum_{k \in \text{MBZ}}[\vAop_{1\sigma  }^{\dagger}(k) , \vBop_{1\sigma  }^{\dagger}(k), \vAop_{2\sigma  }^{\dagger} (k), \vBop_{2\sigma  }^{\dagger}(k)  ]      H_{\sigma }(k) 
	\begin{bmatrix}
		\vAop_{1\sigma } (k)   \\
		\vBop_{1\sigma } (k)\\
		\vAop_{2\sigma  }(k)\\
		\vBop_{2\sigma  }(k)\\
	\end{bmatrix},
\end{align}
where MBZ stands for the $k$-space magnetic Brillouin zone. 
The Hamiltonian matrix for the spin-up electron $H_{\uparrow }(k) $ is given by:
\begin{eqnarray}
	H_{\uparrow}(k)=\left(\begin{array}{cccccc}
		U\langle \vn_{A1\downarrow}\rangle & -t' & -g_k & 0  \\
		-t'& U\langle \vn_{B1\downarrow} \rangle&0 &-g_k \\
		-g_k&0& U\langle \vn_{A2\downarrow}\rangle& -t' \\
		0 &-g_k&-t' & U\langle \vn_{B2\downarrow}\rangle \\
	\end{array}\right), \; g_k = 2 t \cos(k a_0).
	\label{IntTB}
\end{eqnarray}
The matrix indices are in the order of  $1A, \;1B, \;2A, \;2B$, where $A (B)$ refers to the upper (lower) chain sites in a magnetic unit cell {\it}. The dimension of the matrix $H_{\uparrow}$ is four, equal to the number of the atoms in a magnetic unit cell.      
An  eigenstate of $H_{\uparrow}(k)$ is given by a column vector
$(c_{A1 \uparrow}(k), c_{B1 \uparrow}(k), c_{ A2 \uparrow}(k),c_{B2\uparrow}(k))^T$.
The occupation numbers are defined as follows:
\begin{equation}
    \langle \vn_{A1\downarrow}\rangle \equiv \sum_{k < k_F} \vert c_{A1\downarrow}(k) \vert^2, \quad 
     \langle \vn_{B1\downarrow}\rangle \equiv \sum_{k < k_F} \vert c_{B1 \downarrow}(k) \vert^2,
\end{equation}
where $k_F$ is the Fermi wavevector.
Similar expressions can be found for   $\langle \vn_{A2\downarrow}\rangle$   and  $\langle \vn_{B2\downarrow}\rangle $.  
These site occupation numbers should be computed self-consistently from the occupied eigenstates. 
There is another Hamiltonian matrix $H_{\downarrow}(k)$ for spin-down electrons,
which has the same structure as Eq.~(\ref{IntTB}) but with reversed spins.    
In the presence of electron interaction with finite repulsion $U$, the occupation number expectation values $\langle \vn_{mp,\sigma}(l) \rangle$ in real space have periodicity with a length of $2a_0$.

\begin{figure}[hbt!]
\begin{center}
\includegraphics[width = 0.6 \textwidth]{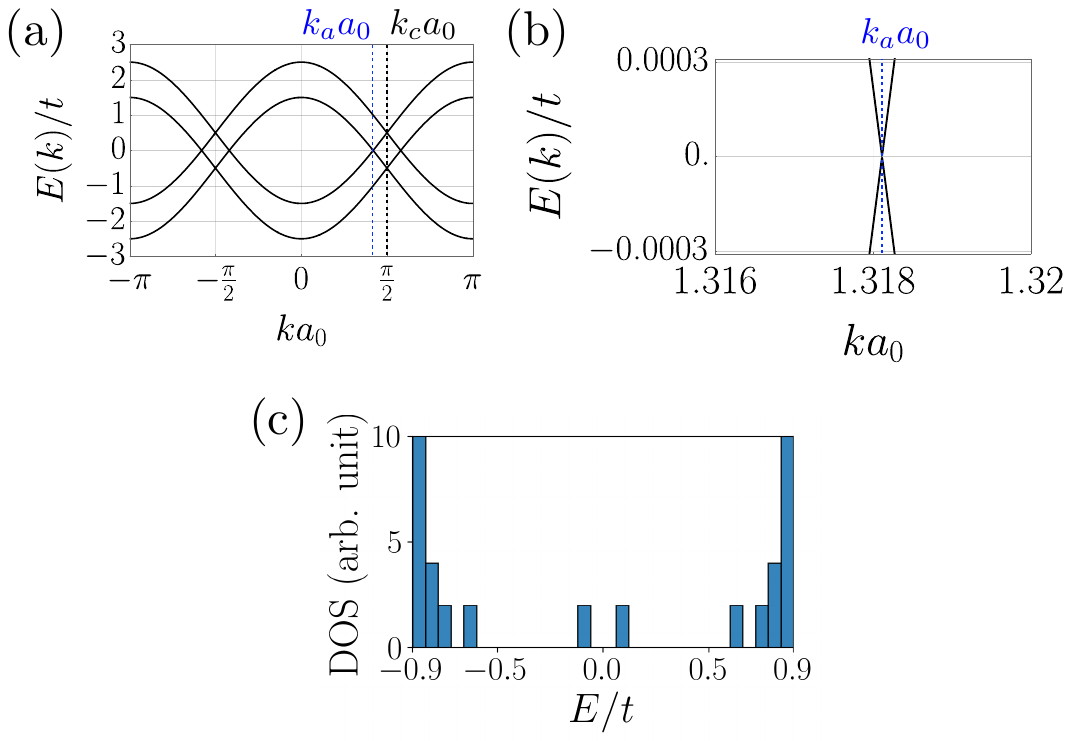}
\caption{(a) The Hartree-Fock band structure for the  parameters  $U=t$ and  $t'=0.5t$. The magnetic zone boundary is at $k_c=\pi/2a_0=k_F$ (black dashed line).
(b) Zoomed-in view of the energy gap near $k_a$ (blue dashed line) and zero energy: the conduction and valence bands anticross at 
 $k_a$.   
(c) Density of states near $E=0$ in the disorder-free and undoped system. The parameters are $(U, \Gamma, t') = (3t, 0, 0.5t)$. The DOS is computed for a system length of $L = 1000 a_0$.}
\label{Band_2Sublattice_2_leg_ladder}
\end{center}	
\end{figure}

Using the mean-field Hamiltonian Eq.~(\ref{twoleg}), we can calculate the self-consistent band structure of the two-leg ladder system with finite interaction strength $U$ as illustrated in Fig.~\ref{Band_2Sublattice_2_leg_ladder}.  
The ground state is \textit{antiferromagnetic}: spins of neighboring sites on opposite chains are antiparallel, in addition to those on the same chain.    
It is important to note that the effective unit-cell length is doubled, which is caused by the periodicity of the antiferromagnetic structure. This enlarged unit cell contains four atoms, which differs from the unit cell with two atoms in the absence of $U$. The enlarged unit cell gives rise to a magnetic Brillouin zone.
Interestingly, the conduction and valence bands anticross each other at 
$k_a \neq k_c$ [$k_c$ is the magnetic zone boundary; see Fig.~\ref{Band_2Sublattice_2_leg_ladder}(a)]. The DOS thus has a gap but is {\it small}, as shown in Fig.~\ref{Band_2Sublattice_2_leg_ladder}(c). The states near $k_c$ and $k_a$
display {\it linear} energy dispersion, namely, Dirac-like dispersion.

\pagebreak

\subsection{Properties of undoped  disordered two-leg ladder}

Now, we introduce disorder into periodic two-leg ladders. We use a diagonal disorder potential $V_i=\sum_{i'} V_{i'}\delta_{i,i'}$, where $V_i\in[-\Gamma,\Gamma]$ represents the random strength of the impurity located at the $i$th lattice site, following a Gaussian distribution.  The following demonstrates the importance of the number of impurities.
 In the continuum limit of the lattice where $V(\vec{r})=\sum_i \gamma_j\delta(\vec{r}-\vec{r}_j)$ and $\gamma_j\in[-\gamma,\gamma]$, 
the variance $\langle V(\vec{r})V(\vec{r}') \rangle$ is given by
 $c\gamma^2\delta(\vec{r}-\vec{r}')$, which depends not only on $\gamma$ but also on the impurity density $c$. We have tested various impurity concentrations.  Our results for fractional charges with an impurity concentration of $n_{\rm imp} = 0.1$ are representative of other values.

\begin{figure}[h!]
\centering
\includegraphics[width=0.5\linewidth]{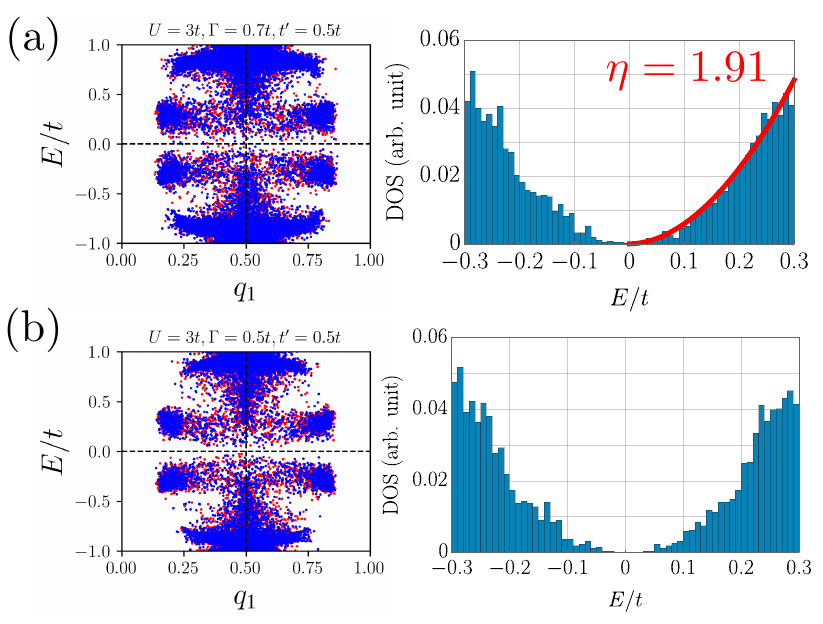}
\caption{The $q_1(E,\uparrow)$ (red dots) and $q_1(E,\downarrow)$ (blue dots), along with the corresponding DOS of several representative cases are plotted with $L = 100a_0$ and 2,000 disorder realizations. The parameters are  (a) $(U, \Gamma, t') = (3t, 0.7t, 0.5t)$ and (b) $(U, \Gamma, t') = (3t, 0.5t, 0.5t)$.    The red line represents a fit to $A(e^{\alpha E^{\eta}}-1)$.}
\label{fig:dos_and_q1_largeGamma}
\end{figure}

We have displayed the DOS in Fig. \ref{fig:dos_and_q1_largeGamma} for two different values of $\Gamma$. In the absence of disorder, the DOS at $E=0$ is finite. However, a significant effect of disorder is the formation of a well-developed {\it soft gap}~\cite{efros1975,YangPRL1993}. This gap is zero at $E = 0$ and gradually increases as $|E|$ increases. It arises from the interplay between disorder and on-site electron repulsion, $U$. As shown in Fig. \ref{fig:dos_and_q1_largeGamma}(a), the numerical soft gap of the DOS can be fitted using $A(e^{\alpha E^{\eta}}-1)$, which represents an exponentially decaying gap near $E=0$.  Such a rapidly decaying gap is important for the stability of the topological order.
  Using the Hartree-Fock eigenstate $\psi_{E,\sigma}(i)$ with site index $i$, we  define 
 \begin{eqnarray}
 q_1(E,\sigma)=\sum_{i\in \text{chain 1}}|\psi_{E,\sigma}(i)|^2.
\end{eqnarray}
It is a measure of the degree of fractionalization in the chain $1$. 
The numerical results of this quantity are shown in Fig.~\ref{fig:dos_and_q1_largeGamma}.
For a fractionalized state $q_1=1/2$, while $q_1=1$ for a localized electron on the chain $1$.
We observe some gap states with $E\approx 0$ and $q_1\approx 1/2$, indicating the presence of well-defined fractional charges. 
(These states represent {\it instanton} states.)

\begin{figure}[hbt!]
\begin{center}
\includegraphics[width = 0.5\textwidth]{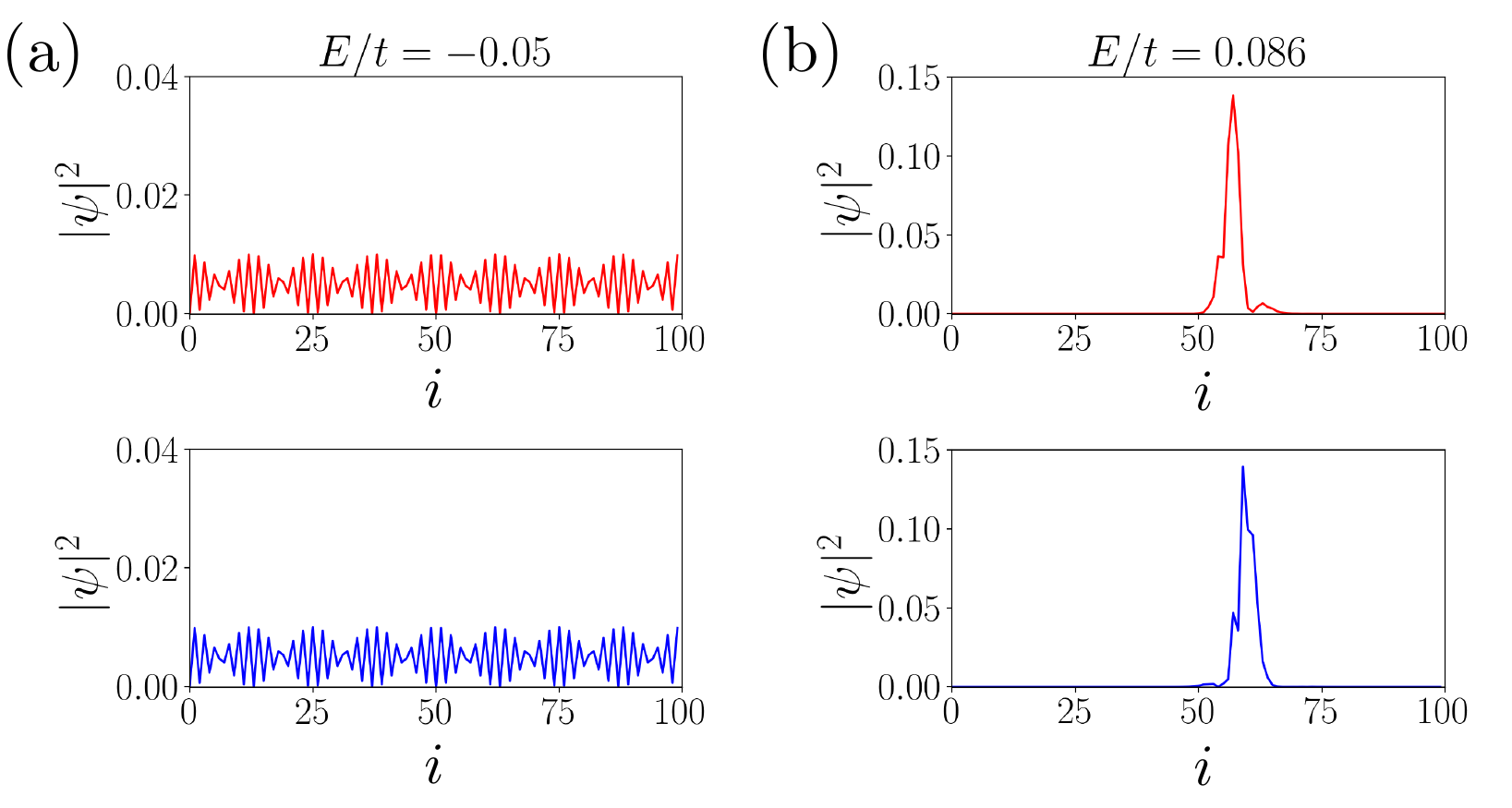}
\caption{ 
(a) The probability density of an extended  state  at the zone boundary $k_c$  in the absence of disorder is shown ($i$ denotes the site index).    Here, the upper (lower) figure shows the probability density at the sites of chain  $m=1  (m=2)$.   The parameters are   $(U, \Gamma, t') = (1.1t, 0, 0.5t)$ and  $L = 100 a_0$ (lattice length).
(b) In the presence of disorder,   {\it localized instanton} states with fractional charges may exist.  An example of such a state  is shown. Our numerical results indicate that well-localized fractional charges—one residing on the upper chain and the other on the lower chain—can emerge with energy $E\approx 0$.
The parameters $(U,\Gamma,t')=(3t,0.5t,0.5t)$ are used with a lattice length of $L = 100 a_0$. 
 }
\label{2edge_frac}
\end{center}	
\end{figure}

Note that the lateral positions of the corresponding fractional charges are nearly {\it identical}, as shown in Fig.~\ref{2edge_frac}(b). For comparison, an extended state of a disorder-free two-leg ladder is shown in Fig.~\ref{2edge_frac}(a).  Electron localization plays an important role in stabilizing fractional charges, as localized fractional states do \textit{not} spatially overlap—a hallmark of localization~\cite{Altshuler, GV2000, yang2022}.
The gap plays an important role in reducing the fluctuations in quantum charge that are neglected in the Hartree-Fock approach~\cite{yang2022}.

\begin{figure}[hbt!]
\begin{center}
\includegraphics[width = 0.7\textwidth]{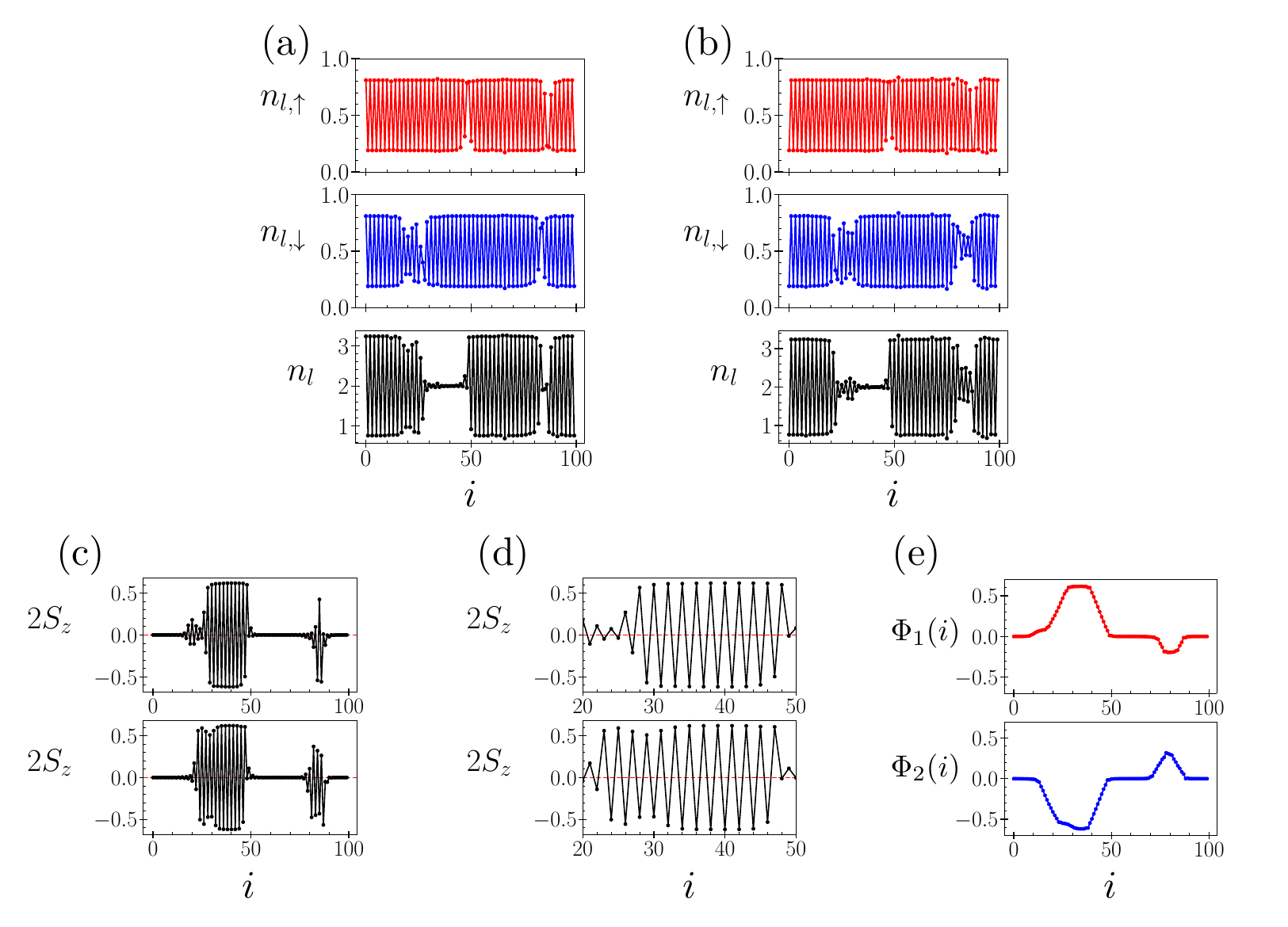}
\caption{The spin-up and spin-down  occupation numbers of chains 1 (a) and 2 (b), along with their site spins,  are shown for a lattice size of  $L = 100 a_0$. The parameters $(U,\Gamma,t')=(3t,0.5t, 0.5t)$ are used. Figure (a) corresponds to chain 1, and Fig. (b) corresponds to chain 2.
(c-d) Site spin values, along with a zoomed-in view of each chain are shown for a lattice size of $L = 100 a_0$. Upper (lower) figures correspond to chain 1 (2). Spins of neighboring sites on opposite chains are antiparallel, in addition to those on the same chain.
(e) Local magnetic order parameters $\Phi_1$ and $\Phi_2$ for chain 1 and 2  are plotted. Upper (lower) figures correspond to chain 1 (2).}
\label{2edge_ground}
\end{center}	
\end{figure}

Fractional charges create magnetic kinks in the order parameter. 
Figures~\ref{2edge_ground}(a) and \ref{2edge_ground}(b) show significant variation in spin-up and spin-down occupation numbers along the length of the disordered ladder.  The site spins, $\langle \operator{S}_z\rangle=\left( \langle \vn_{i\uparrow}\rangle-\langle \vn_{i\downarrow}\rangle \right)/2$,
are shown for each chain in Figs. \ref{2edge_ground}(c) and \ref{2edge_ground}(d). Figure~\ref{2edge_ground}(d) reveals that neighboring spins from different chains are antiferromagnetically coupled, implying that the order parameters have opposite signs, as illustrated in Fig. \ref{2edge_ground}(e). 
Note that the fractional charges of an instanton reside where the local antiferromagnetic order parameters $\Phi_{1,2}(i)$ change rapidly, and their lateral positions on chains 1 and 2 are {\it nearly identical}. In our study, an instanton thus exhibits a special positional correlation between chains 1 and 2.

\pagebreak

\subsection{Topological entanglement entropy}

The presence of gap states with fractional charges strongly suggests that two-leg ladders are topologically ordered insulators. If this is the case, the topological entanglement entropy must be nonzero and universal, independent of the system parameters.

\begin{figure}[hbt!]
\begin{center}
    \includegraphics[width = 0.25\textwidth]{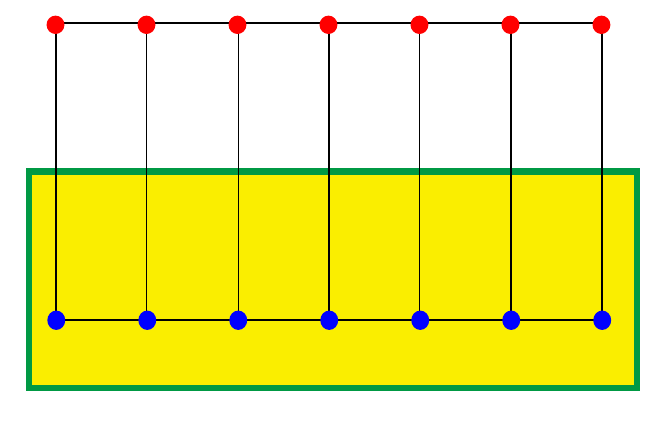}
    \caption{A two-leg ladder is divided into two equal regions.  The domain $D$ is highlighted in yellow.}
    \label{twolegladder_region}
\end{center}	
\end{figure}

We now proceed to calculate the topological entanglement entropy~\cite{Balents2012,Wen2006,Kitaev2006}. The entanglement entropy $S_D=-\text{Tr}[\rho_D \text{ln}\rho_D]$  is given by the reduced density matrix $\rho_D$ of region $D$.
 A  simply connected region $D$ in a topologically ordered  gapful system has the following expression for the entanglement entropy in the limit as the system length $L\rightarrow \infty$: 
\begin{equation}
    S_D = \alpha L - \beta.
    \label{TopoEntropy}
\end{equation}
 The first term is owing to local correlations, and $\alpha$ is a nonuniversal constant. The second term $\beta$ represents the subdominant and universal topological entanglement entropy originating from non-local correlations.
The entanglement entropy  is computed as 
\begin{eqnarray}
S_D=-\sum_i
[\lambda_i \ln \lambda_i-(1-\lambda_i)\ln (1-\lambda_i)],
\end{eqnarray}
where the $\lambda_i$ are the eigenvalues of the following correlation function (or, equivalently, of the reduced density matrix of region $D$, as shown in Ref.~\cite{yang2021}):
\begin{eqnarray}
C_{\vec{R},\vec{R}'}=\langle \Psi| \vcop^+_{\vec{R}\uparrow} \vcop_{\vec{R}'\uparrow}|\Psi\rangle.
\label{Corr}
\end{eqnarray}
In Eq.~(\ref{Corr}), $\vec{R}, \vec{R}'\in D$ denote positions and $\vert \Psi \ra$ is the ground state.
We choose the region $D$ so that it includes the lower chain sites, as shown in Fig. \ref{twolegladder_region}. This choice is equally effective when using the correlation function of spin-down states and for the upper chain sites. 

The Wilson loop method~\cite{Wen2006} is preferable for computing the parameter $\beta$ because of its computational efficiency, as it requires fewer disorder realizations (typically $N_D < 100$), particularly in systems such as zigzag graphene nanoribbons~\cite{yang2021}.
 However, this method cannot be applied to two-leg ladder systems, as the geometry is too narrow to support the formation of a Wilson loop. Instead, for these systems, we divided the ladder into two halves, as illustrated in Fig.~\ref{twolegladder_region}. This approach required a significantly larger number of disorder realizations. To ensure convergence, we computed \( \beta \) while gradually increasing \( N_D \) (e.g., \( N_D = 300,\; 600,\; 900,\; 1200, \ldots \)) until the variance in \( \beta \) was sufficiently reduced.
The largest system length we used was $L\sim 10000 a_0$.

\begin{figure}[h!]
\centering
\includegraphics[width=0.6\linewidth]{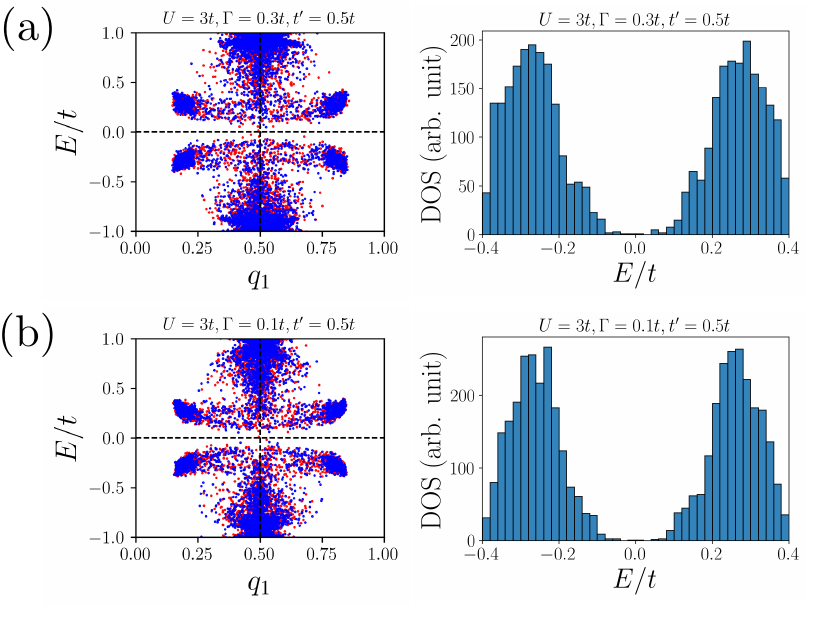}
\caption{The $q_1$-$E$ diagram and the corresponding DOS of several representative cases are plotted with $L=100a_0$ 
 and 2,000 disorder realizations. The parameters are  (a) $(U, \Gamma, t') = (3t, 0.3t, 0.5t)$ and (b) $(U, \Gamma, t') = (3t, 0.1t, 0.5t)$.}
\label{fig:dos_and_q1_smallGamma}
\end{figure}

For an accurate calculation of $\beta$, the correlation length must be short, which can be calculated from the correlation function
Eq.~(\ref{Corr}). We choose the parameters such that the correlation length \cite{Pachos_2012} along the transverse direction is nearly zero.
For the parameters $U = 3t$, $t' = 0.5t$, and $\Gamma = 0.5t$, the average value of the correlation function at one site of a chain is 0.5, while at the opposite site it is significantly reduced to 0.084. For a smaller $t'=0.2t$, this value decreases further to 0.03. Thus, smaller values of $t'$ are desirable to reduce the correlation length. However, when $t'$ is too small, the chains become independent of each other.
 It is not uncommon for some topologically ordered systems to have very short correlation lengths; for example, toric code systems have a correlation length of zero~\cite{Kitaev2003,Zeng2015}.
For smaller disorder strengths, a nearly hard gap (which does not follow an exponential form) persists instead of a soft gap, as shown in Fig.~\ref{fig:dos_and_q1_smallGamma}. The desirable values of $\Gamma$ should be much smaller than $U$ 
but not excessively small, ensuring a well-formed soft gap (hereafter referred to as {\it the soft gap regime}). Otherwise, $S_D$ displays significant variance, and as a result,  $\beta$ becomes
 non-universal.

\pagebreak

\begin{figure}[hbt!]
\begin{center}
    \includegraphics[width = 0.4\textwidth]{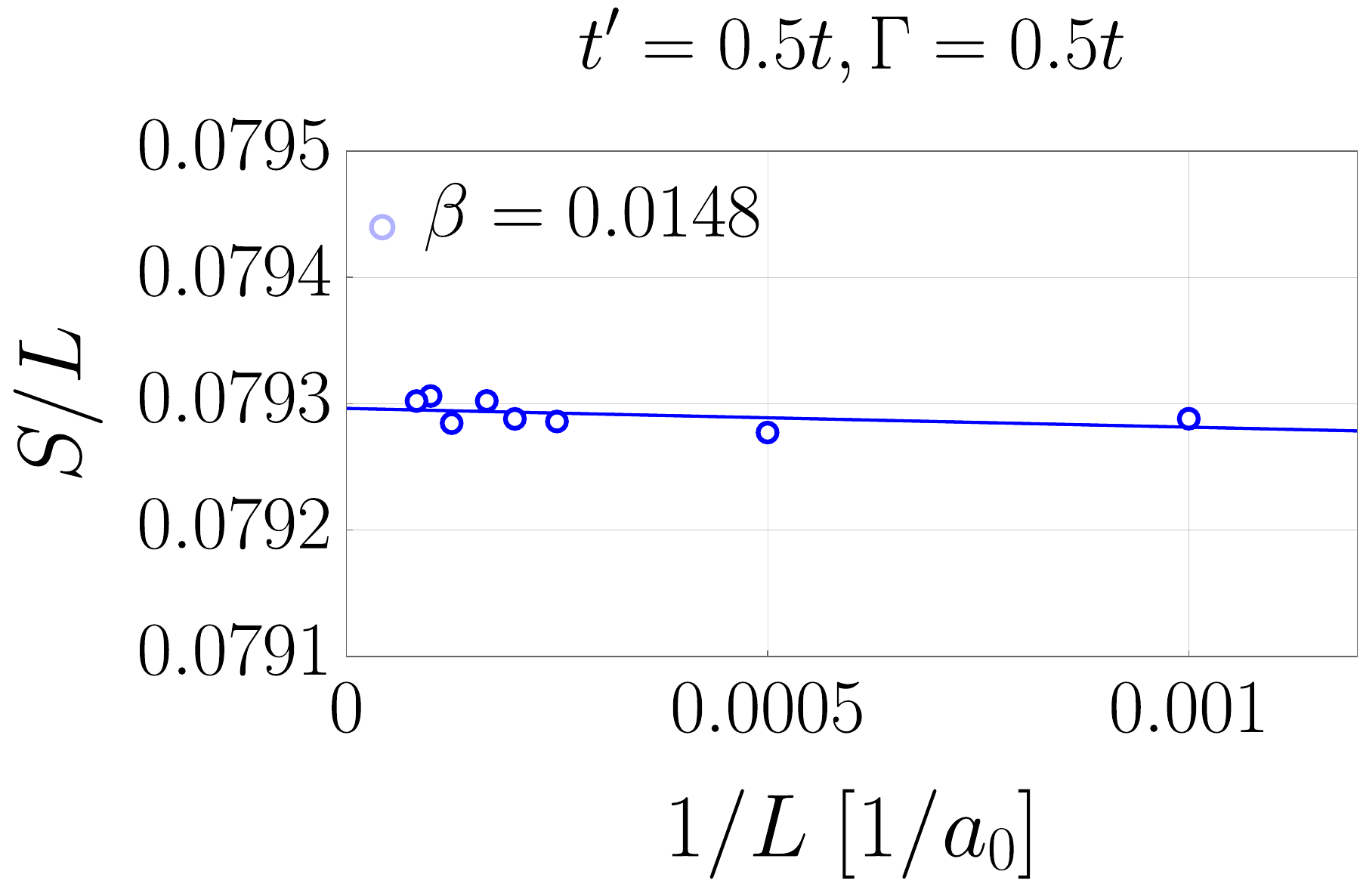}
    \caption{The average entanglement entropy per site of the partitioned region $S/L$ is computed as a function $1/L$. The parameters are $U=3t$,  $t'=0.5t$, $n_{\rm imp}=0.1$, and $\Gamma=0.5t$. For each point the number of disorder realizations is larger or equal 400. The slope of the fitted line is the negative of the topological entanglement entropy $-\beta$.}
    \label{EntEntro1}
\end{center}	
\end{figure}

\begin{figure}[hbt!]
    \centering
    \includegraphics[width=0.4\linewidth]{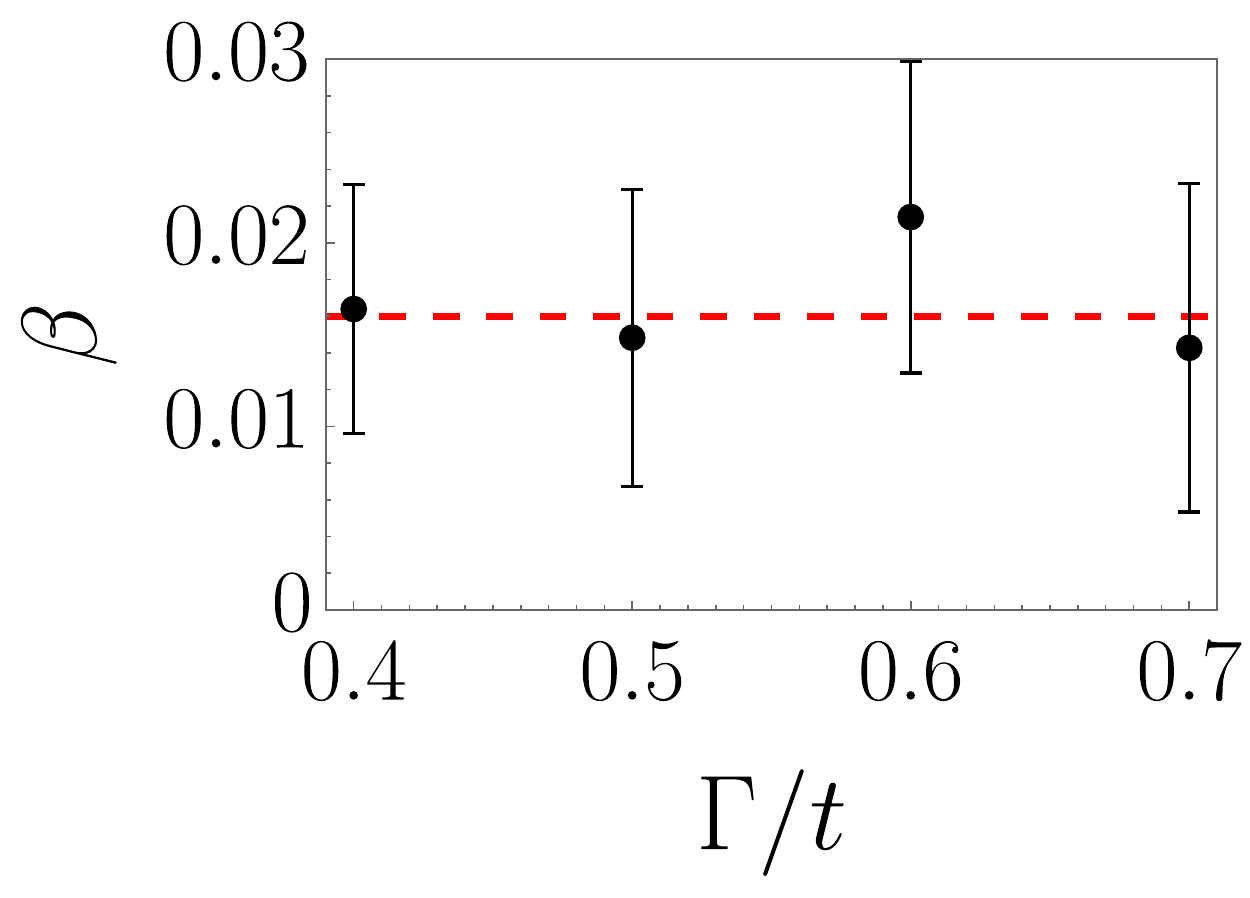}
    \caption{Plots of $\beta$ as a function of $\Gamma/t$ for $(U, t') = (3t, 0.5t)$. Red-dashed line indicates the universal value of topological entanglement entropy of ZGNRs, $\beta \approx 0.016$.   }
    \label{universalBeta}
\end{figure}

A finite scaling analysis is used to compute   the Hartree-Fock  topological entanglement  entropy~\cite{yang2021}  of a two-leg ladder.
\textcolor{black}{Using Eq.~(\ref{TopoEntropy}), the quantity $S_D/L$ can be plotted against $1/L$, and the slope of the resulting linear fit yields $-\beta$}, as shown in Fig.~\ref{EntEntro1}. (We performed GPU calculations on a superconductor to
accelerate our computations.)
A systematic investigation of the parameter space $(U/t, \Gamma/t, t'/t)$ reveals a specific region where $\beta\approx 0.016$ is universal, independent of $\Gamma$, $t'$, and $U$.   An example is shown in Fig.~\ref{universalBeta}. 
{\it This is precisely the region in which a soft gap is well developed.}
Outside this region, the variance of entanglement entropy $S_D$
begins to increase, and only quasi-topological order starts to emerge~\cite{Le_2024}.  When the disorder becomes sufficiently strong, the topological order vanishes, as magnetism, the gap, and the kinks in the order parameter are all destroyed.

\begin{figure}[hbt!]
	\begin{center}
		\includegraphics[width = 0.6\textwidth]{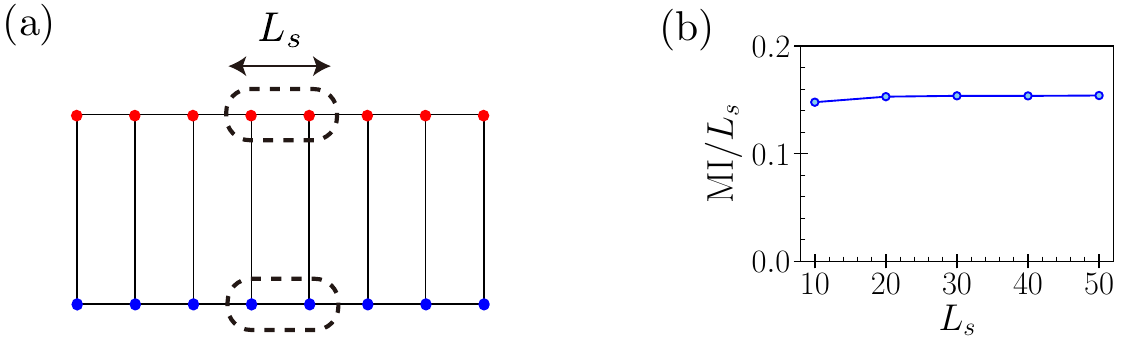}
		\caption{(a) A two-leg ladder is divided into two regions: one region ($I$) consists of the sites within the dashed ovals on one chain, while the other region ($J$) comprises the sites on the opposite chain.  (b)  The mutual information $M_{I,J}$ between  the oval regions, divided by the length of an oval size $L_s$, is displayed.  The parameters used for this figure are $(U, \Gamma, t') = (3t, 0.5t, 0.5t)$.}
		\label{EntEntro2}
	\end{center}	
\end{figure}

In quantum information, the mutual information is used to analyze correlations in quantum systems, which can be expressed in terms of entropy.
We have computed the Hartree-Fock mutual information~\cite{mut_inf2023} between two regions $I$ and $J$, defined as
	\begin{eqnarray}
	M_{I,J}=S_{I}+S_{I}-S_{I,J}.
\end{eqnarray}
Here, $S_{I}$, $S_{J}$, and $S_{I,J}$ denote the entanglement entropy of regions $I$, $J$, and their union $I \cup J$, respectively.  \textcolor{black}{We define region \( I \) (\( J \))---a segment of length \( L_s \)---as a subregion of chain 1 (2), as illustrated in Fig.~\ref{EntEntro2}(a).}
Our calculation shows that this mutual information is {\it finite} between chains 1 and 2. For sufficiently large regions, it is proportional to their length, as shown in Fig.~\ref{EntEntro2}(b). We believe that the entanglement entropy of fractional charges between chains 1 and 2 contributes to mutual information.

\pagebreak

\section{Shankar-Witten-type approach} 
\label{section:Bosonization1D}

So far, we have used a numerical approach to explore the properties of two-leg ladders. We now turn to a bosonized Shankar-Witten-type Lagrangian, which introduces key concepts and provides insight into our numerical findings. This framework offers a more compact formulation for understanding the effects of disorder on the magnetic order parameter, as well as for generating kinks and instantons.

\subsection{A brief introduction to bosonization method of one-chain model}
\label{sec:bosonization}

Before we present the bosonized Lagrangian of disordered systems, we briefly review the basic elements of the bosonization method. Bosonization is a method that can express low-energy (near Fermi energy) fermion operators in terms of density modulation and its dual (both of them are bosonic, hence the name bosonization).
There are diverse conventions for the bosonization formula, and the following formulas explicitly indicate our convention.  An exhaustive treatment of the bosonization method can be found in Refs.~\cite{Gogolin, Giamarchi}, and we follow the convention of Ref.~\cite{Delft1998} regarding the treatment of zero modes (see below).

The non-interacting one-dimensional low-energy fermions have right and left moving modes with 
linear dispersion, and they constitute a Dirac fermion:
\begin{equation}
\widehat{\psi} = (\widehat{\psi}_R, \widehat{\psi}_L).
\label{DiracW} 
\end{equation}
Then the original fermion operator $\widehat{\vc}(x)$ near Fermi energy can be written as ($k_F$ is Fermi momentum)
\begin{equation}
\label{chiral-decompose}
 \vcop(x) \sim  e^{i k_F x}  \widehat{\psi}_R(x) +  e^{-i k_F x}  \widehat{\psi}_L(x),
\end{equation}
where the higher harmonics terms involving $e^{\pm i n k_F x}$ with $n \ge 2$ 
are neglected (here $x$ labels the  position along the chain direction, and $k_F=k_c$).

From the perspective of the density profile, fermion creation and annihilation correspond to a kink and an antikink, respectively~\cite{Giamarchi}.
 The boson operator that describes the kink is denoted by $\widehat{\theta}$. 
Additionally, to account for the anticommutation relations, a phase factor is required, which is described by the \textit{dual} boson operator $\widehat{\phi}$.
Now the bosonization formula of a single species of fermion takes the following form:
\begin{equation}
\widehat{\psi}_R(x) = \frac{\widehat{F}_R}{\sqrt{2\pi a}} \, e^{+i \widehat{\theta}(x)+i \widehat{\phi}(x)}, \quad 
\widehat{\psi}_L(x) = \frac{\widehat{F}_L}{\sqrt{2\pi a}} \, e^{-i \widehat{\theta}(x)+i \widehat{\phi}(x)},
\label{Boson}
\end{equation}
where a phase factor that becomes small for large system size is neglected for each formula.
$a$ is a short distance cutoff (of the order of lattice spacing), and
$\widehat{F}_{R/L}$ are Klein factors that are made up of zero modes (see below),  and they are necessary for the anti-commutation between {\color{black}$\widehat{\psi}_R$ and $\widehat{\psi}_L$}.
Two boson field operators $\widehat{\theta},\widehat{\phi}$ are dual to each other in the following sense:
\begin{equation}
\label{dual-commutator}
 [\widehat{\phi}(x), \widehat{\theta}(x')] = i \frac{\pi}{2} \mathrm{sign}(x-x').
\end{equation}

The normal-ordered density operators of chiral fermions $\widehat{\psi}_{R/L}$ are given by ($L_\text{sys}$ is a system size)
\begin{equation}
\widehat{\rho}_R(x)=:\widehat{\psi}_R^\dag(x) \widehat{\psi}_R(x): =\frac{1}{2\pi} \frac{\partial \widehat{\phi}_R(x)}{\partial x}+\frac{\widehat{N}_R}{L_\text{sys}}, \quad
\widehat{\rho}_L(x)=:\widehat{\psi}_L^\dag(x) \widehat{\psi}_L(x): =\frac{1}{2\pi} \frac{\partial \widehat{\phi}_L(x)}{\partial x}+\frac{\widehat{N}_L}{L_\text{sys}},
\end{equation}
where the chiral bosons are defined by (the definition varies depending on conventions)
\begin{equation}
\widehat{\phi}_{R}=\widehat{\theta} +\widehat{\phi}, \quad \widehat{\phi}_{L}=\widehat{\theta} -\widehat{\phi},
\end{equation}
and $\widehat{N}_{R/L}$ is the number of extra fermions added on vacuum state (of filled Dirac sea) (these are the zero modes mentioned above, and see the discussions in Sec.~\ref{sec:semion-comm}). 
The operators $\widehat{\theta}$ and $\widehat{\phi}$ do not include zero modes in our convention~Ref. \cite{Delft1998}.
Then the net density operator is given by
\begin{equation}
\label{density-operator}
\widehat{\rho}(x) = \widehat{\rho}_R(x) + \widehat{\rho}_L(x) = \frac{1}{\pi} \frac{\partial \widehat{\theta}}{\partial x}+\frac{\widehat{N}_R+\widehat{N}_L}{L_\text{sys}}.
\end{equation}
If we implicitly assume that
the density includes the contribution away from the Fermi point,
the last term of Eq.~(\ref{density-operator}) can be regarded as the average electron density.

\subsection{Shankar-Witten-type Lagrangian of one-chain model}
\label{sec:SW-onechain}

It is well known that the one-dimensional Hubbard model can be mapped to a Luttinger liquid at low energies if
$U/t $ is not too large. In this subsection, we write down a Shankar-Witten-type bosonized Lagrangian~\cite{Shankar1978} for disordered and interacting one-dimensional systems.

The (real time) \textit{path integral Lagrangian} of the continuum {\it one-chain} model in a disorder potential $V(x)$ (in both fermionic Grassmann  and bosonized variables)
is given by the following: 
\begin{eqnarray}
\mathcal{L}_\text{tot}&=&	\mathcal{L}_\text{el} +	\mathcal{L}_m +\mathcal{L}_\text{dis},\nonumber\\
\mathcal{L}_\text{el} &=& \int dx \Big[ 
\psi^\dag_L(i \p_t -i v_F \p_x) \psi_L + \psi^\dag_R(i \p_t +i v_F \p_x) \psi_R
-g\Phi (\psi^\dag_R \psi_L + \psi^\dag_L \psi_R ) \Big ]  \nonumber \\
&=&
\frac{1}{2\pi} \int dx \Big ( \frac{1}{v_F} (\partial_t \theta)^2-v_F (\partial_x \theta)^2 - g \Phi \cos (2 \theta) \Big ), \nonumber\\
\mathcal{L}_m &=& \int dx  \Big[\frac{1}{2 v_0^2} (\p_t \Phi)^2- \frac{1}{2}  (\partial_x \Phi)^2-\lambda(\Phi^2-\tilde{a}^2)^2 \Big ],\nonumber\\
\mathcal{L}_\text{dis}&=&\int dx  V(x) \left(\rho_R+\rho_L +e^{-2 i k_F x} \psi^\dag_R \psi_L + e^{2 i k_F x} \psi^\dag_L \psi_R \right) \nonumber \\
&=&\int dx \left( V_0(x) \rho(x)+ (V_{2 k_F} (x)  \psi^\dag_R \psi_L  +\text{h.c.}) \right),
\label{Lag}
\end{eqnarray}
where  the path integral variables are denoted without a hat to indicate that they are no longer operators.  $\mathcal{L}_\text{el} $ describes the low-energy Dirac electron (in both fermionic and bosonized forms) and its coupling to the magnetization order parameter $\Phi$, which is determined self-consistently from the occupied electronic states. Here, $g$ is the corresponding coupling constant, $k_F$ denotes the Fermi wavevector, and  $v_F$ represents the Fermi velocity.
The third line of Eq.~(\ref{Lag}) is obtained using the bosonization formula in Eq.~(\ref{Boson}) and by rewriting 
\begin{equation}
	\psi^\dag_R \psi_L + \psi^\dag_L \psi_R  \sim  \cos (2 \theta),
\end{equation}
which is now expressed in terms of path integral variables.  Here, the Klein factors (involving zero-modes only) are neglected. $\mathcal{L}_{m}$ represents the Ginzburg-Landau Lagrangian \cite{Goldenfeld} for the $2k_F$ order parameter, which is invariant under the transformation $\Phi(x) \rightarrow -\Phi(x)$.  $v_0$ denotes the characteristic velocity of the order parameter, $\lambda$ is the coupling constant for the quartic self-interaction of the order parameter, and $\tilde{a}$ determines the (bare) minimum of the quartic potential of $\Phi$.
The last term, $\mathcal{L}_\text{dis}$, represents the electron-disorder interaction, where $V_0(x)$ corresponds to components with momenta near zero, and $V_{\pm 2k_F}(x)$ corresponds to components near the momentum $\pm 2k_F$. Specifically, $V_{0}(x) = \int_{-\Lambda}^{+\Lambda} d q \; e^{-i q x} V_q$ and $V_{\pm 2k_F}(x) = \int_{\pm 2k_F - \Lambda}^{\pm 2k_F + \Lambda} d q \; e^{-i q x} V_q$, 
where $\Lambda$ is an appropriate cutoff (of the same order as the cutoff for the validity of Dirac linear dispersion).
For a given disorder realization $V(x)$, $V_{\pm 2k_F}(x)$ pins the charge density profile $\theta(x)$ at specific positions $x_i$ via backscattering, while $V_0(x)$ is irrelevant for pinning.
For a model with a dilute impurity concentration, we can effectively replace  $\mathcal{L}_{\rm dis}$ with
\be
\mathcal{L}_{\text{dis}} \longrightarrow \sum_i  \big ( t_i \psi^\dag_R(x_i) \psi_L (x_i) + \mathrm{h.c.} \big ),
\label{disEff}
\ee
where $t_i$ represents the effective pinning strength.
The total Lagrangian $\mathcal{L}_\text{tot}$ then represents an \textit{effective bosonization} theory. It is valid for a spin-up or spin-down one-chain system in the limit of small $U$ (with spin indices suppressed).

To understand the main physics described by the  Lagrangian Eq.~(\ref{Lag}), it is useful to first consider the system in the absence of disorder. From the Lagrangian $\mathcal{L}_\text{el}+\mathcal{L}_{m}$, one can search for the static classical solutions \textit{that interpolate the minima of effective potentials} $g \Phi \cos (2 \theta) + \lambda (\Phi^2 -\tilde{a}^2)$~\cite{Shankar1978} (see Fig.~\ref{Potential_Witten_a=1,lambda=1,g=1}).  
They will represent the excitations of one-chain system upon quantization.
There are two kinds of minima~\cite{Shankar1978}:
$( \theta = n \pi, \Phi = -\tilde{a}')$ and $(\theta=(n+\frac{1}{2}) \pi, \Phi= +\tilde{a}')$ ($n$ is an integer). 
Here, $\tilde{a}' \neq \tilde{a}$ owing to the coupling $g \Phi \cos 2 \theta$. 
The time-independent classical solution connecting  the adjacent minima with fixed $\Phi$ describes the original fermion operator with fermion number 1 (see Eq. (\ref{density-operator}))
\begin{equation}
 \Delta N = \frac{1}{\pi} \int dx \frac{\partial \theta}{\partial x} = \frac{\Delta \theta}{\pi} = \pm 1.
\end{equation}
However, the solution interpolating between two adjacent minima with \textit{different} values of $\Phi$ 
[an example of such a solution is the kink, as shown in Fig.~\ref{quantum_chain_ground}(b)]
yields
\begin{equation}
 \Delta N = \frac{\Delta \theta}{\pi} = \pm \frac{1}{2},
\end{equation}
which corresponds to an electronic soliton state residing at the position of the kink.
This state is a midgap electronic state that is only half-filled \cite{HeegerMod1988}.

\begin{figure}[hbt!]
\begin{center}
    \includegraphics[width = 0.4\textwidth]{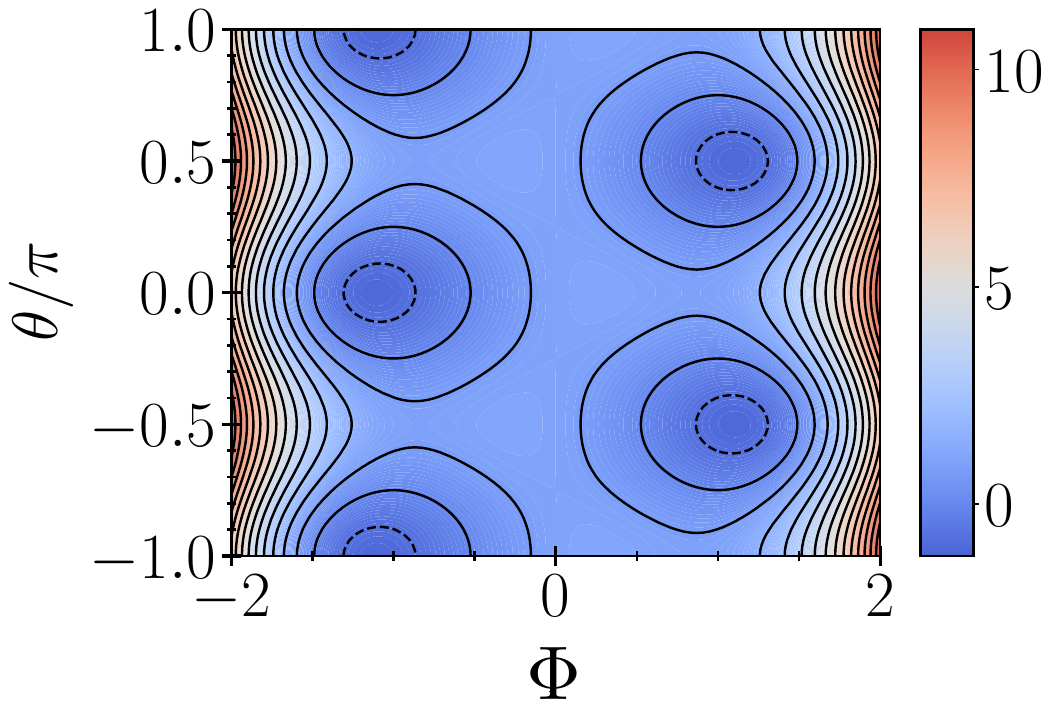}
    \caption{The effective potential of the one-chain model without disorder.   The potential has with minima  at 
    $(\Phi=-1, \theta = n \pi)$  and at $(\Phi=1, \theta= (n+\frac{1}{2})\pi)$. The parameters are $\lambda=1, \tilde{a}=1$, and $g=1$.}
    \label{Potential_Witten_a=1,lambda=1,g=1}
\end{center}	
\end{figure}

Now, let us try to understand the numerical results from Section \ref{section:TwoLeg} in the presence of disorder, from the perspective of the above semiclassical treatment of the effective bosonization model. We expect that disorder will couple the ground state and excited states with kinks. Indeed, in the presence of (weak) disorder, our numerical results show that the order parameter can develop kinks at multiple sites [see Fig.~\ref{2edge_ground}(e)].  In particular, we find that the site occupation number is $n_l=1$ in the absence of disorder, whereas in the presence of disorder, it increases to approximately $n_l \approx 1.5$. 
In terms of the Shankar-Witten-type Lagrangian of the one-chain model, this means that the order parameter $\Phi$ can develop kinks, positioned at specific locations $x_i$, accompanied by an electronic soliton state \cite{Su1979, HeegerMod1988, Jackiw1976}.
The quantities $(\Phi(x), g, \{x_i\})$ can be determined by judiciously choosing from the numerically computed electronic states for a given disorder potential $V(x)$ (see Sec. \ref{section:TwoLeg}).  Or, one could minimize the total energy.
The disorder breaks the chain into several regions with different signs of $\Phi(x)$, and a soliton state connects the two different regions.  


\subsection{Shankar-Witten-type Lagrangian of two-leg ladder in the presence of disorder}
\label{section:Shankar-Witten-type Lagrangian}

\begin{figure}[hbt!]
\begin{center}
\includegraphics[width = 0.4\textwidth]{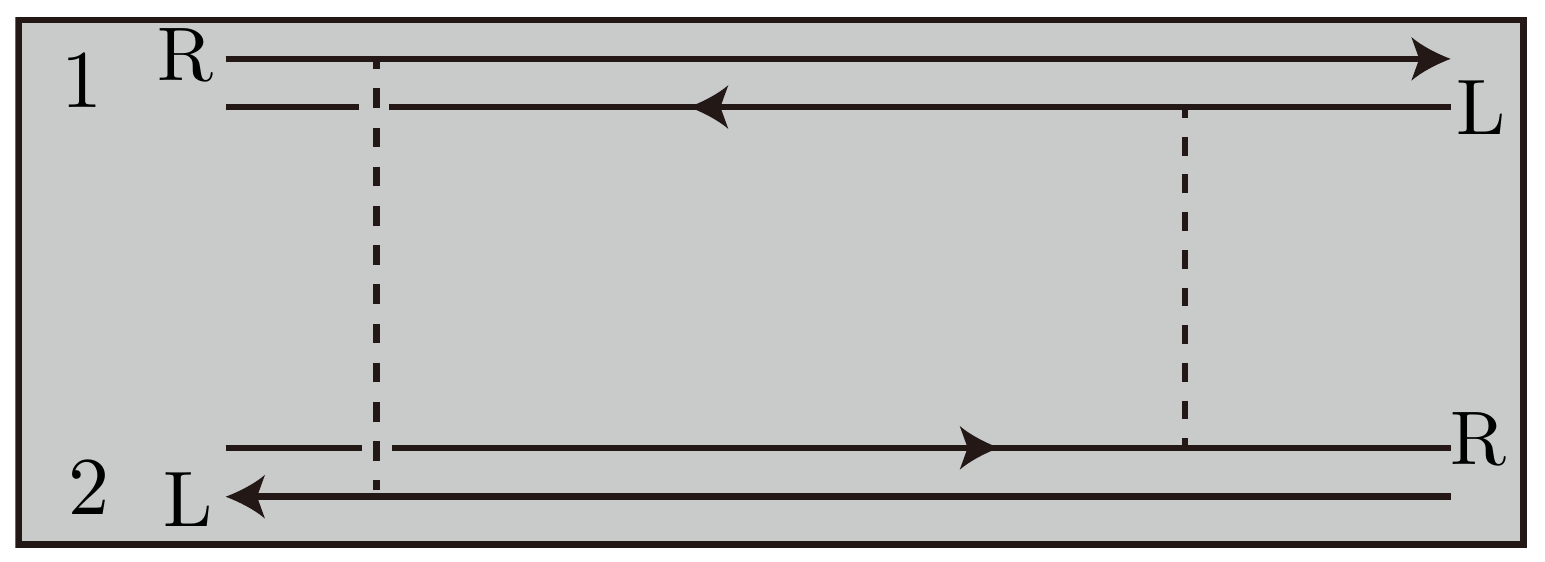}
\caption{In a continuum model of a two-leg ladder, the right- and left-moving states of chains 1 and 2 are shown schematically.  Dashed lines represent inter-chain coupling  between the left and right movers,  induced by tunneling and/or disorder. 
Schematic band structure of a  disorder-free but interacting system is shown in Fig. \ref{Band_2Sublattice_2_leg_ladder}.}     
\label{edges1}
\end{center}	
\end{figure}

A two-leg ladder can also be thought of as consisting of two coupled chains.
Utilizing the bosonization method for a single chain, as described in 
Sec.~\ref{sec:SW-onechain}, we present a bosonized Lagrangian for disordered and interacting two-leg ladder systems. 
For each chain and spin component, there are right ($R$) and left ($L$) movers, as illustrated in Fig. \ref{edges1}. Thus, there are four types of fermions, denoted by $L1, L2, R1$, and $R2$.
The two chains are coupled by both a tunneling term and disorder.
This inter-chain interaction will create a singular perturbation and generate an instanton.


The following Lagrangian for a two-leg ladder in a continuum model is obtained by generalizing the one-chain Lagrangian discussed in
Sec.~\ref{sec:SW-onechain}:
\begin{align}
&		\mathcal{L}_{\rm two-leg}=\mathcal{L} _1+\mathcal{L} _2+	\mathcal{L} _{\rm inter}+\mathcal{L}_\text{dis},\nonumber\\
&	\mathcal{L} _1= \int dx \Big[ \frac{1}{2} ( \sum_{\mu=t,x} \partial_\mu \Phi_{1})^2+\psi^\dag_{1L}(i \p_t -i v_F \p_x) \psi_{1L} + \psi^\dag_{1R}(i \p_t +i v_F \p_x) \psi_{1R}-g\Phi_{1} (\psi^\dag_{1R} \psi_{1L}+\text{H.c.}) +V(\Phi_1)
\Big ],\nonumber\\
&	\mathcal{L} _2=\int dx \Big[ \frac{1}{2} ( \sum_{\mu=t,x} \partial_\mu \Phi_{2})^2+\psi^\dag_{2L}(i \p_t -i v_F \p_x) \psi_{2L} + \psi^\dag_{2R}(i \p_t +i v_F \p_x) \psi_{2R}-g\Phi_{2} (\psi^\dag_{2R} \psi_{L2}+\text{H.c.})+V(\Phi_2)
\Big ],\nonumber\\
& \mathcal{L} _{\rm inter}= \int dx  \, t_{\rm inter}(x) \big( \psi^{\dag}_{1R}(x)\psi_{2L}(x)+\text{H.c.} \big),\nonumber\\
&\mathcal{L}_\text{dis}=
\int dx   \left( U_1(x)e^{-2 i k_F x} \psi^\dag_{1R} \psi_{2L} 
+U_2(x) e^{2 i k_F x} \psi^\dag_{1L} \psi_{2R }+\text{H.c.} \right)
+\int dx  V_1(x) (\rho_{1R}+\rho_{1L} )+\int dx  V_2(x) (\rho_{2R}+\rho_{2L} ).
\nonumber\\\label{Lag0}
\end{align}
Here $\psi_{1L}(x)$, $\psi_{1R}(x)$, $\psi_{2L}(x)$, and  $\psi_{2R}(x)$ denote the  Dirac electrons in chains $1$ and $2$. $\Phi_{1}(x)$ and $\Phi_{2}(x)$ denote the magnetization order parameters of chains $1$ and $2$.  
$V(\Phi_{i=1,2})$ is the quartic potential of $\mathcal{L}_{m=1,2}$ in Eq. (\ref{Lag}), $\mathcal{L} _{\rm inter}$ is the interchain coupling, and $\mathcal{L}_\text{dis}$ is the disorder potential.
In one-dimensional systems, the dominant effect of disorder arises 
from the back-scattering term (the first term 
of $\mathcal{L}_\text{dis}$).

\begin{figure}[hbt!]
	\begin{center}
		\includegraphics[width = 0.3\textwidth]{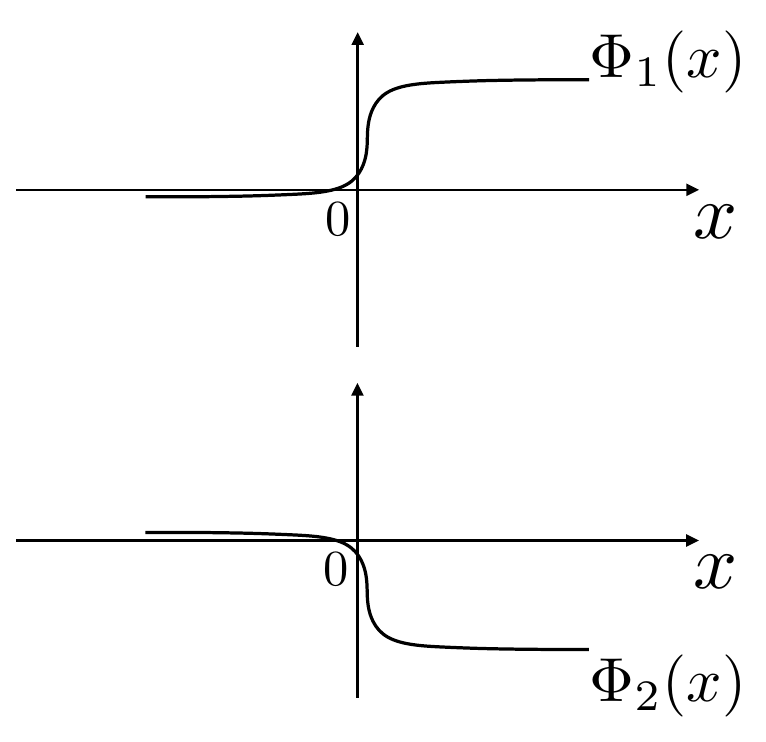}
		\caption{ Schematic drawing of antiferromagnetic order parameters of chains 1 and 2 with kinks.   }
		\label{edgeOrder}
	\end{center}	
\end{figure}

The above model can be reduced to a simpler one that qualitatively describes instantons in disordered two-leg ladders. The effects of disorder and interchain hopping effectively generate instantons at positions $x_i$, which can be simulated, in the limit of   dilute impurity concentration, by replacing
 $\mathcal{L} _{\rm inter}+\mathcal{L}_\text{dis}$ with
\be
\mathcal{L}_{\rm coupling }= \sum_i t_i \big( \psi^{\dag}_{1R}(x_i)\psi_{2L}(x_i)+\psi^{\dag}_{2R}(x_i)\psi_{1L}(x_i)+\text{H.c.} \big)\\,
\ee
where $t_i$ represents the interchain coupling at position $x_i$, induced by inter-chain hopping and disorder. In this simplified model, we impose the constraint that the opposite edges are \textit{antiferromagnetically} coupled (see Fig. \ref{edgeOrder}): 
\be \Phi_{1}(x) = -\Phi_{2}(x).
\ee
This simplified model represents the numerical results shown in Fig. \ref{2edge_ground}(e) fairly well.
For example, when an electron fractionalizes into an instanton at $x_i=0$, the order parameters develop kinks on both chains and change sign on opposite chains, as shown in Fig. \ref{edgeOrder}.
 In our model,  both $x_i$ and $t_i$ can be fitted to the numerical results. We can also determine the quantities $\Phi_{1,2}$ and the coupling constant $g$ from the numerical data.


\pagebreak
It should be noted that the formation of an instanton is not a sufficient condition for stable fractional charges, as quantum fluctuations can destabilize them~\cite{Jackiw1983}. When disorder is not too strong, localization stabilizes fractional charges by suppressing quantum fluctuations, thereby reducing variance in the entanglement entropy~\cite{Le_2024}. Moreover, the presence of an excitation gap further mitigates this effect~\cite{GV2000}.\\

\section{Fusion of Semions in Topologically Ordered Two-Leg Electron Ladder}
\label{section:Bosonization2Leg}
In this section, we propose a bosonization description of the fusion of $e/2$ fractionally charged semions, which serves as a low-energy effective field theory. This approach may offer deeper insight into instantons~\cite{yang2020} in two-leg ladders. Additionally, it provides a qualitative understanding of the numerical results through approximate analytical methods.
\subsection{Motivation for studying instantons}
\label{subsection:semion1}
An instanton consists of two $e/2$ fractional charges.   These charges are expected to be {\it semions}, meaning they exhibit intermediate statistical behavior between fermions and bosons.
How can their existence be tested?  The following heuristic argument suggests a useful approach. 
Suppose an electron tunnels into a two-leg ladder and fractionalizes into two   non-overlapping  fractional charges as depicted in Fig. \ref{2edge_frac}(b),
\begin{eqnarray}	
	e\rightarrow e/2+e/2.
\end{eqnarray}
In this picture, no gap exists in the energy spectrum. The resulting    tunneling density of states is expected to be linear: The 
$I\mbox{-}V$ curve is given by~\cite{GV2019}
\begin{eqnarray}
	I\propto\int d\epsilon_1 \int d\epsilon_2 \, H_s (eV-\epsilon_1 -\epsilon_2)\propto V^2, 
	\label{IV}
\end{eqnarray}
where
$H_s$ and $\epsilon_{1,2}$ are the Heaviside step function and quasiparticle energies, respectively. This $I\mbox{-}V$ curve is equivalent to a linear tunneling DOS.  A linear DOS is thus a good indication of the presence of fractional charges.

\begin{figure}[hbt!]
\begin{center}
\includegraphics[width = 0.5\textwidth]{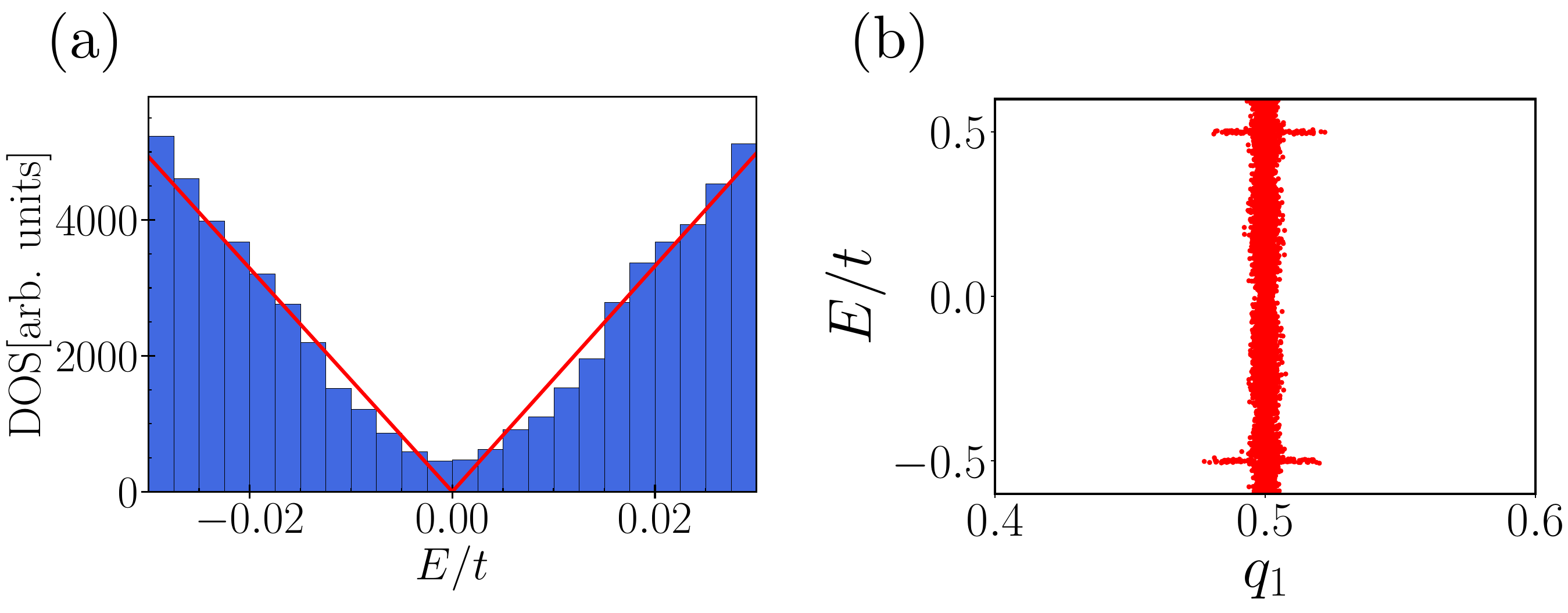}
\caption{(a) Approximately linear  DOS  for $(U, t', \Gamma) = (t, 0.5t, 0.45t)$, with 3,000 disorder realizations. The red line is a linear fit. States become less localized as $\vert E \vert$ increases from $0$. (b) $(q_1,E)$ plot for the identical parameters to (a).}
\label{qA1}
\end{center}	
\end{figure}

Is this result consistent with our numerical result?  At a certain disorder strength, our numerical results indeed show a linear DOS, as shown in Fig.~\ref{qA1}(a).
To connect the linear gap to the possible presence of $e/2$ fractional charges, we have also investigated the probability that an electron is located on chain 1: $q_1=\sum_{i\in \text{chain  1}}   |\psi(x_i)|^2$ [see Fig.~\ref{qA1}(b)]. The condition $q_1=1/2$ is necessary for charge fractionalization. We observe that  $q_1\approx 1/2$ 
for {\it most eigenstates} [in the absence of disorder, states are delocalized, as shown in Fig.\ref{2edge_frac} (a)].      
Numerical results show that the Hartree-Fock eigenstates are dominated by instantons consisting of two fractional charges.


\subsection{Instanton as a pair of semions}
\label{sec:semion-comm}

The discussion in Sec.~\ref{subsection:semion1} suggests the presence of fractional charges.
A fractional charge is highly correlated and nonlocal object from the electron point of view. 
Is it mathematically consistent to consider an instanton as a pair of semions?  It depends on how it is constructed.

In our approach, described below, we construct an instanton (electron) consisting of one semion residing on chain 1, with the other residing on chain 2, as depicted in Fig.~\ref{exsem1}.
 Such an object turns out to be mathematically consistent, as will be shown below.   Our numerical results also indicate that {\it an electron is unlikely to break up into two on the same chain}, as evidenced by the probability $q_1$
 of the electron being localized on chain 1,  which is centered around $1/2$ for low-energy states ($E\sim 0$), as shown in Figs.~\ref{fig:dos_and_q1_largeGamma} and \ref{qA1}(b).  

\begin{figure}[hbt!]
	\begin{center}
		\includegraphics[width = 0.25\textwidth]{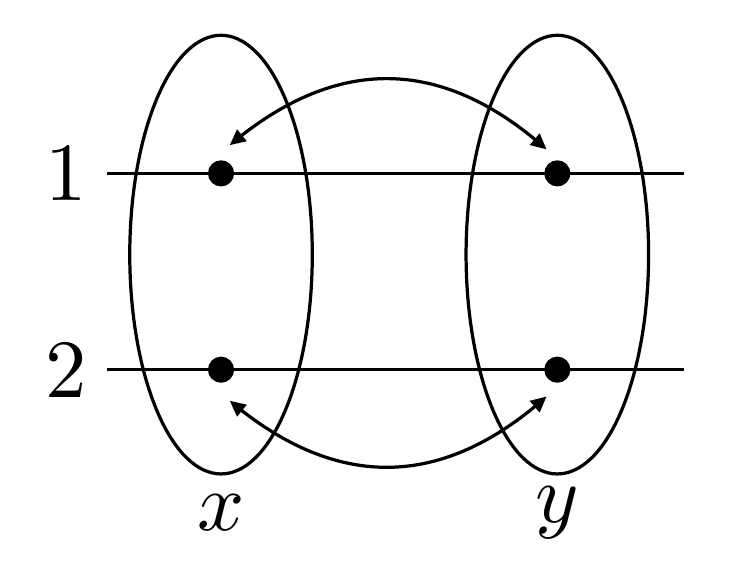}
		\caption{Black dots represent semions, and an electron is shown inside an oval. One semion can reside on chain 1 while the other resides on chain 2.
        The arrows indicate semion exchanges.} 
	\label{exsem1}
	\end{center}	
\end{figure}

In our model, the {\it semion operators} are defined in terms of bosonic phase fields and Klein factors as follows.
The left-moving semion destruction operator of  $j$th chain ($a$ is a short distance cutoff) is given by
\begin{equation}
\label{quasiL}
\widehat{\psi}_{s,j L}(x) = \frac{1}{\sqrt{2\pi a}}\, \widehat{F}_{s,j L} \, 
 e^{-i \widehat{\phi}_{j L}},
 \end{equation}
 while that of the  right-moving semion operator is given by
\begin{equation}
\label{quasiR}
\widehat{\psi}_{s,j R}(x) = \frac{1}{\sqrt{2\pi a}}\, \widehat{F}_{s,j R} \, 
 e^{+i \widehat{\phi}_{j R}},
 \end{equation} 
where an unimportant phase factor that becomes smaller for large system size is neglected.  $\widehat{F}_{s,j R/L}$ is the Klein factors for semion quasiparticles (not for electrons) that are to be constructed from the zero-modes only. In the above, the boson operators $\widehat{\phi}_{R/L,j}$ include only the non zero-mode oscillators \cite{Delft1998}. These will be explained in more detail below.

\subsection{Semion phase fields}

 \textcolor{black}{To construct an appropriate effective Lagrangian for the semion phase fields, we draw an analogy with the edge Lagrangian of fractional quantum Hall states~\cite{Laughlin1983}. For odd-denominator filling factors $\nu = \frac{1}{3}, \frac{1}{5}, \ldots$, the following real-time Lagrangian~\cite{Wen1992} captures the phase fields associated with the chiral edge excitations,
 \begin{eqnarray}
	\mathcal{L}_{\rm semion}&=&\frac{1}{4\pi\nu}\{ -\partial_x\phi_{1,R}(\partial_t+v_F\partial_x)\phi_{1,R}- \partial_x\phi_{1,L}(-\partial_t+v_F\partial_x)\phi_{1,L}
	\} \nonumber\\
	&+&\frac{1}{4\pi\nu}\{-\partial_x\phi_{2,R} (	\partial_t+v_F\partial_x)\phi_{2,R}-  \partial_x\phi_{2,L} (-\partial_t+v_F\partial_x)\phi_{2,L}\}. \nonumber\\
	\label{Lag1}
\end{eqnarray}
However, at filling factor $\nu = 1/2$, the physics changes: An electron (instanton) cannot be constructed from two semions on the {\it same} chain, since such an object would not satisfy the fermionic anti-commutation relations (a semion and an anti-semion can fuse to form a boson on the same edge~\cite{Haldane1991, Kalmeyer1987}).  Instead, an electron can be formed when the boson combines with a neutral fermion. Consequently, Eq.~(\ref{Lag1}) does not provide an adequate description of two-leg ladders. Nevertheless, with appropriate modifications, the Lagrangian can be adapted to describe semions localized separately on chains 1 and 2, such that a semion from chain 1 and a semion from chain 2 together form an electron, as explained below.}

 \textcolor{black}{We adopt Eq.~(\ref{Lag1}) as an effective bosonic Lagrangian for the semion phase fields,} which describe spin-up and spin-down semions carrying an even-denominator fractional charge (with spin indices suppressed). In this formulation, left- and right-moving semions reside on chains 1 and 2, and their corresponding chiral boson phase fields are denoted by $\phi_{i,R/L}$~\cite{Yang1996}, where $i=1,2$ labels the chains and $R/L$ denotes right- and left-moving bosons, respectively.
Note the sign difference between the right and left movers.   We assume that this Lagrangian is quantized using the following commutation relations for chiral bosons, where classical boson fields are promoted to operators,
\begin{align}
	\label{chiral-comm}
	[\widehat{\phi}_{i,R}(x), \widehat{\phi}_{j,R}(x') ] &= + i \pi \, \nu \,\text{sign}(x-x')\, \delta_{ij},  \quad 
 [\widehat{\phi}_{i,L}(x), \widehat{\phi}_{j,L}(x') ] = - i \pi \, \nu \,\text{sign}(x-x')\, \delta_{ij},  \nonumber \\
 [\widehat{\phi}_{i,R}(x), \widehat{\phi}_{j,L}(x') ] & = 0,  \quad  i,j = 1,2.
\end{align}
These commutation relations are \textit{formally} identical to those of odd-denominator edge states, except that here, the crucial factor 
$\nu=1/2$ on the right-hand side gives rise to the semionic character.
Note that the commutator between the 1 and 2 bosons vanishes.  However, we will introduce the Klein factors below, which do not commute [see Eq.~(\ref{klein2})].
   In our approach, the semion itself is constructed as an operator involving these chiral boson phase fields and Klein factors.  An instanton (electron) then consists of a semion residing on chain 1 and another on chain 2, rather than both residing on the same chain, which would lead to an inconsistency.
 Below, we will show that the Lagrangian 
$\mathcal{L}_{\rm semion}$ and the commutation relations yield a physically consistent description of instantons.

\textcolor{black}{The assumptions underlying the proposed Lagrangian, along with the accompanying commutation relations, are as follows: (A) A disorder term must be added to the Lagrangian in Eq.~(33). It should not, however, be strong enough to make the DOS finite at $E=0$. At the same time, it should not be so weak that a hard gap remains. (B) The Lagrangian serves as a low-energy effective theory, derived via bosonization, which is valid in the low-energy regime with Dirac-like linear-energy dispersion in the absence of disorder (see Fig.~\ref{Band_2Sublattice_2_leg_ladder}). In the presence of disorder, the bosonization approach remains applicable only if the energy cutoff of the low-energy regime is greater than, or at least comparable to, the disorder strength. (C) The interchain coupling $t'$ must not be too small; otherwise, well-defined semions on individual chains are unlikely to emerge. In this limit, the chains behave nearly independently, resembling disordered, interacting one-dimensional systems.}

\subsection{Construction of semion Klein factors}

How do we construct  these Klein factors?   In the usual bosonization formula of \textit{electrons},
the Klein factors are necessary for the anticommutation
relation between different species of chiral \textit{electron} operators.  We choose the corresponding 
Klein factor for the semion quasiparticle, denoted as  $\widehat{F}_{s,i L/R }$, in the following form \{these formulas were motivated by Eq.~(8.2.22) of Ref.~\cite{Polchinski}\}:
\begin{align}
\label{klein-q}
 \widehat{F}_{s,1 L} &= e^{i\pi (\widehat{N}_{1L}+\widehat{N}_{2L}+\widehat{N}_{1R}+\widehat{N}_{2R})/2 }\, e^{-i \widehat{\phi}_{0,1L}}, \quad 
  \widehat{F}_{s,2 L} = e^{-i\pi (\widehat{N}_{ 1L}+\widehat{N}_{2L}+\widehat{N}_{1R}+\widehat{N}_{2R})/2}\, e^{-i \widehat{\phi}_{0,2L}},
  \nonumber \\
 \widehat{F}_{s,1 R} &= e^{i\pi (\widehat{N}_{1R}+\widehat{N}_{2R}+\widehat{N}_{1L}+\widehat{N}_{2L})/2 }\, e^{+i \widehat{\phi}_{0,1R}}, \quad  
  \widehat{F}_{s,2 R} = e^{-i\pi (\widehat{N}_{1R}+\widehat{N}_{2R}-\widehat{N}_{1L}-\widehat{N}_{2L})/2 }\, e^{+i \widehat{\phi}_{0,2R}}
\end{align}
where $\widehat{\phi}_{0,i L/R}$ is the left(right)-moving zero mode of boson satisfying (note that the semion nature is reflected in the factor $\nu=1/2$ below)
\begin{equation}
\label{klein1}
[\widehat{\phi}_{0,i L}, \widehat{N}_{j L }] = +i \nu \delta_{ij}, \quad [\widehat{\phi}_{0,i R}, \widehat{N}_{j R }] = +i \nu \delta_{ij},
\end{equation}
where $\widehat{N}_{j L/R}$ is the conjugate zero mode of $\widehat{\phi}_{0,i L/R}$.
We comment on the technical aspects of the zero modes of a one-dimensional boson 
(for detailed accounts, see pp.~238--239 of~\cite{Polchinski} and pp.~28--30 of~\cite{Gogolin}).
The boson fields $\theta, \phi$ in the bosonization formalism are compact (i.e., periodic variables  like angular variables) since
they should be exponentiated to represent fermion.  Additionally, for a finite system (of size $L_s$) with certain boundary conditions (e.g., periodic boundary conditions), nontrivial zero modes arise.  These zero modes can be decomposed into the left- and right-moving modes (consider a single species). These modes  consist of a constant term, $\widehat{\phi}_{0 L}$ and  $\widehat{\phi}_{0 R}$, and a linear function of space and time, $\widehat{N}_L (x + v t)/L_s$ and  $\widehat{N}_R(x-v t)/L_s$, as free bosons satisfy the second-order wave equation (of velocity $v$).
 Then the imposition of  the canonical
boson commutation relations leads to Eq.~(\ref{klein1}).
Now $\widehat{N}_{1 L/R}$ and $ \widehat{N}_{2 L/R}$ are interpreted to be the number operator for the  left(right)-moving semions in the chain 1 and 2, respectively.

In the present formulation, the zero modes are assigned to the Klein factors. The boson operators $\widehat{\phi}_{j R/L}$ are understood to include only the nonzero modes, so that they commute with the zero mode operators
$\widehat{\phi}_{0,j L}$ and $\widehat{\phi}_{0,j R}$, as well as the Klein factors, by construction.
Direct computations (for left-moving semions), using the relation
\( e^A e^B = e^B e^A  e^{[A,B]},\)
which holds if $[A,B]$ is a commuting scalar, show that
\begin{equation}
\label{klein2}
 [\widehat{N}_{i L}, \widehat{F}_{s,j L}] = - \nu \widehat{F}_{s,j L} \delta_{ij}, \quad 
 \widehat{F}_{s, 1 L } \widehat{F}_{s, 2 L } = e^{-i \pi \nu} \widehat{F}_{s, 2 L } \widehat{F}_{s,1 L}.
\end{equation}

\subsection{Semion and electron exchanges}
First, we must demonstrate that the exchange relation between the semion operators $\widehat{\psi}_{s,1L}(x)$ and $ \widehat{\psi}_{s,1 L}(y) $ has the correct form.
They have the same Klein factors that commute with the boson
operators $\widehat{\phi}_{1 L}(x)$ and $\widehat{\phi}_{1 L}(y)$.
Again using the operator relation
\( e^A e^B = e^B e^A  e^{[A,B]},\)
which holds if $[A,B]$ is a commuting scalar, 
we can show that 
\begin{equation}
	e^{-i \widehat{\phi}_{1L}(x)} e^{-i \widehat{\phi}_{1L}(y)} 
	= e^{-i \widehat{\phi}_{1L}(y)} e^{-i \widehat{\phi}_{1L}(x)}  e^{ +i \pi \nu 
		\mathrm{sign}(x-y)}.
\end{equation}
The above result with $\nu=1/2$ leads to the \textit{expected} correct semion statistics
\be
\widehat{\psi}_{s, 1 L }(x) \widehat{\psi}_{s, 1 L}(y)
=e^{ +i \pi \nu \, \mathrm{sign}(x-y)} \widehat{\psi}_{s,1 L }(y) \widehat{\psi}_{s,1 L}(x).
\ee

If an electron is viewed as a composite of two semions residing on different chains (see Fig.~\ref{exsem1}), then the canonical fermion anti-commutation relation must hold for such electrons.
We construct the electron operator explicitly.  It is assumed that an electron fractionalizes into two delocalized and independent semions at the 1 and 2 edges, so that the phase boson fields of different chains  \textit{commute}.
We then define the (left-moving  and right-moving) \textit{electron} operator as the composite semion operator:
\begin{equation}
	\widehat{\Psi}_{e,L}(x) := \widehat{\psi}_{s,1L}(x) \widehat{\psi}_{s,2L}(x), \quad 
    \widehat{\Psi}_{e,R}(x) := \widehat{\psi}_{s,1R}(x) \widehat{\psi}_{s,2R}(x).
    \label{composite-e}
\end{equation}
If the above assumption of fractionalization is correct, 
these composite operators must satisfy the ordinary electron anti-commutation relation.
We can check the anti-commutation relation of $\widehat{\Psi}_{e,L}(x)$ and $ \widehat{\Psi}_{e,L}(y) $ directly
using Eqs.~(\ref{klein-q})--(\ref{klein2}). Unlike the semion exchange, the Klein factors are now crucial because semions from different chains are exchanged.
First, consider 
\( \widehat{\psi}_{s, 2 L }(x) \widehat{\psi}_{s,1 L}(y) \) and try to commute them. Remember that two left-moving boson operators $\widehat{\phi}_{1 L}$ and 
$\widehat{\phi}_{2 L}$ commute each other. 
Thanks to the relation
\begin{equation}
\widehat{F}_{s,1 L } \widehat{F}_{s, 2L  } = e^{-i \pi \nu} \widehat{F}_{s,2 L} \widehat{F}_{s,1 L},
\quad \widehat{F}_{s,2 L } \widehat{F}_{s,1 L} = e^{+i \pi \nu} \widehat{F}_{s,1 L} \widehat{F}_{s,2 L},
\end{equation}
we obtain
\begin{align}
\widehat{\psi}_{s, 2L }(x) \widehat{\psi}_{s, 1 L }(y)
&=\widehat{F}_{ s,2L } e^{-i \widehat{\phi}_{2 L}(x) }\,\widehat{F}_{s,1 L} e^{-i \widehat{\phi}_{1 L}(y) } \nonumber \\
&=\widehat{F}_{s,2 L}\widehat{F}_{s,1 L} e^{-i \widehat{\phi}_{1 L}(y) }e^{-i \widehat{\phi}_{2 L}(x) }
=e^{+i \pi \nu} \widehat{\psi}_{s,1 L }(y) \widehat{\psi}_{s,2 L}(x).
\end{align}
Using the above results we can find the commutation 
relation of between two composite left-moving electron operators $\widehat{\Psi}_{e,L}(x)$ 
and $\widehat{\Psi}_{e,L}(y)$ as follows:
\begin{align}
\widehat{\Psi}_{e,L}(x) \widehat{\Psi}_{e,L}(y)&=
\widehat{\psi}_{s,1L}(x) \widehat{\psi}_{s,2L}(x) \widehat{\psi}_{s,1L}(y) 
\widehat{\psi}_{s,2L}(y) \nonumber \\
&=e^{+i \pi \nu} \widehat{\psi}_{s,1L}(x)  \widehat{\psi}_{s,1L}(y) \widehat{\psi}_{s,2L}(x) \widehat{\psi}_{s,2L}(y) \nonumber \\
&=e^{+i \pi \nu} e^{ +2 i \pi \nu \mathrm{sign}(x-y)}
\widehat{\psi}_{s,1L}(y)  \widehat{\psi}_{s,1L}(x) \widehat{\psi}_{s,2L}(y) \widehat{\psi}_{s,2L}(x)  \nonumber\\
&=e^{+i \pi \nu} e^{ +2 i \pi \nu \mathrm{sign}(x-y)} e^{-i \pi \nu}
\widehat{\psi}_{s,1L}(y) \widehat{\psi}_{s,2L}(y) \widehat{\psi}_{s,1L}(x)  \widehat{\psi}_{s,2L}(x)  \nonumber \\
&= e^{ +2 i \pi \nu \mathrm{sign}(x-y)} \widehat{\Psi}_{e,L}(y) \widehat{\Psi}_{e,L}(x), \quad \nu =1/2, \\
&=(-1) \widehat{\Psi}_{e,L}(y) \widehat{\Psi}_{e,L}(x).
\end{align}
We indeed find that the electron operator $\widehat{\Psi}_{e,L}(x)$ satisfies the ordinary anti-commutation relation of fermion (electron) as expected.
We can also check that the anti-commutation relation between 
$\widehat{\Psi}_{e,L}(x)$ and $\widehat{\Psi}_{e,R}(y)$, but the calculations (of Klein factors) are much more tedious.


\section{Calculation of the DOS based on the CDW pinning model}
\label{sec:cdw}

Our numerical results in Fig.~\ref{fig:dos_and_q1_largeGamma} demonstrate that the DOS develops a soft gap near $E=0$.  Moreover, at a critical disorder strength, a linear gap may emerge. We present a calculation of the DOS near 
$E=0$, intended only for qualitative insight rather than a precise numerical comparison.  It uses the instanton (electron) operator that we constructed in the preceding subsections using the bosonization method. We utilize the CDW pinning \textcolor{black}{model~\cite{Fukuyama1978, *larkin1979pinning}}, which provides a qualitative description of the effects of disorder on Luttinger liquids; see Ref.~\cite{Giamarchi} and references therein.
(In the pinning models, the CDW phase is treated as a classical field. In contrast, bosonization expresses fermionic operators in terms of quantum bosonic phase fields, incorporating the effects of quantum fluctuations.)

In our approach, an electron is considered a composite object consisting of two semions on different chains, as shown in Fig.~\ref{exsem1}. This is because instanton physics plays a crucial role. 
This observation led to our proposed definition Eq.~(\ref{composite-e}) of electron operator. 
 The Klein factors that involve zero modes only are not important in dynamical correlation function and we will express
the electron operators in terms of chiral bosons $\phi_{iL}, \phi_{iR}$ only~\cite{Yang1996} (note the chiral boson fields of different chain commute):
\be
\label{composite-e2}
\widehat{\Psi}_{e,R} \sim e^{+i \widehat{\phi}_{1R} +i \widehat{\phi}_{2R}}, \quad 
\widehat{\Psi}_{e,L} \sim e^{-i \widehat{\phi}_{1L} -i \widehat{\phi}_{2L}}.
\ee
It should be emphasized that the electron (instanton) operators of Eq.~(\ref{composite-e2}) are different from
the electron operators $\widehat{\psi}_{1/2,L/R}$ defined in each chain of Eq.~(\ref{Lag0}).
Now let us define phase boson operators,
\be
\label{tilde-boson}
\widehat{\tilde{\theta}} = \frac{1}{2} \big( \widehat{\phi}_{1R} + \widehat{\phi}_{2R} + \widehat{\phi}_{1L} + \widehat{\phi}_{2L} 
\big ), \quad 
\widehat{\tilde{\phi}} = \frac{1}{2} \big( \widehat{\phi}_{1R} + \widehat{\phi}_{2R} - \widehat{\phi}_{1L} - \widehat{\phi}_{2L} 
\big ).
\ee
It should be noted that the tilde boson operators of Eq.~(\ref{tilde-boson}) are different from those in Sec.~ \ref{sec:bosonization}
since here the bosons $\widehat{\phi}_{iL}$ and $\widehat{\phi}_{iR}$ represent \textit{semions}.
In terms of boson fields of Eq.~(\ref{tilde-boson}), the electron operators of Eq.~(\ref{composite-e2}) can be re-expressed as 
\be
\label{composite-e3}
\widehat{\Psi}_{e,R} \sim e^{+i \widehat{\tilde{\theta}} + i \widehat{\tilde{\phi}} }, \quad 
\widehat{\Psi}_{e,L} \sim e^{-i \widehat{\tilde{\theta}} + i \widehat{\tilde{\phi}} }.
\ee

To construct an effective Lagrangian for the disordered system, 
we first note that Eq.~(\ref{tilde-boson}) involves only the symmetric combination of two chains.
So we extract the symmetric part of the semion Lagrangian Eq.~(\ref{Lag1}) and express it in terms of the tilde boson fields of Eq.~(\ref{tilde-boson}). In terms of Hamiltonian, it takes the following form:
\be
H_{0} = \frac{v_F}{4\pi \nu} \int dx \left[ (\p_x \tilde{\theta})^2 + (\p_x \tilde{\phi})^2 \right].
\ee
In one-dimensional system, the dominant effect of disorder comes from the back-scattering term [see Eq.~(\ref{Lag})] that can be written as 
$ \int dx \,(
\tilde{V}_{2 k_F}(x) \widehat{\Psi}_{e,R}^\dag(x) \widehat{\Psi}_{e,L}(x) +\text{H.c.})$,
where $\tilde{V}_{2 k_F}(x)$ is the $2 k_F$ component of the \textit{effective} one-dimensional disorder potential experienced   by the \textit{composite} electron operator.
The back-scattering term takes the following form:
\be
\label{semion-disorder}
\widehat{H}_{\rm bs} = \int dx V_{\rm dis} \cos (2 \widehat{\tilde{\theta}}+ \zeta),
\ee
 where we represent the backscattering part of the effective disorder potential as
$ \tilde{V}_{2 k_F}(x) = V_{\rm dis}(x) e^{i\zeta(x)}$ with a random phase $\zeta(x)$. Note that $\tilde{V}_{2 k_F}(x)$ is generally complex.

It is well known that the effect of disorder in a one-dimensional system can be understood from the perspective of the pinning of the charge density wave.  In our notation, the $\tilde{\theta}(x,\tau)$ phase determines the local density profile. 
For a given random phase of disorder in Eq.~(\ref{semion-disorder}), $\tilde{\theta}(x,\tau)$ is adjusted to a static $\tilde{\theta}_0(x)$ to optimize the energy. (For a  discussion on this point, see Ref.~\cite{Giamarchi}, and we will use the imaginary time $\tau$ in this subsection for convenience.)  
The $\tilde{\theta}(x,\tau)$ phase is thus pinned within the localization length $L_0$, which corresponds to the pinning length.  Next, within the segment, small fluctuations $\delta\tilde{\theta}(x,\tau)$ near this optimal $\tilde{\theta}_0(x)$ are included: The potential for cosine disorder is expanded and parameterized as $V_0 (\delta\tilde{\theta(x,\tau)})^2/2$, where $V_0$ represents the strength of a typical disorder potential well, which also depends on the localization length \cite{Giamarchi}. Note that the local current, given by the spatial derivative of $\tilde{\phi}$, may be non-zero, but the total current is zero because a pinned charge density wave is an insulator.

Once the $\tilde{\theta}(x,\tau)$ phase boson is nearly pinned, its dual boson $\tilde{\phi}(x,\tau)$ locally fluctuates wildly, eventually dominating the local electron Green's function [$\tilde{\theta}$ and $\tilde{\phi}$   are conjugate variables (see Eq.~\ref{dual-commutator}), analogous to the position and momentum $x$ and $p$]. 
In the bosonization formula,  Eq.~(\ref{composite-e3}),
the right-moving local electron Green's function  can be approximated as
\begin{align}
	G_{e,R}(x,x, \tau)&  = - \la \widehat{\Psi}_{e,R}(x,\tau)  \widehat{\Psi}^\dag_{e,R}(x,0) \ra
    =-\la e^{i (\tilde{\theta}_0(x) + \delta \widehat{ \tilde{\theta}}(x,\tau)) + i \widehat{\tilde{\phi}}(x,\tau)}  
    e^{-i (\tilde{\theta}_0(x) + \delta \widehat{\tilde{\theta}}(x,0))-i  \widehat{\tilde{\phi}}(x,0)} \ra \nonumber \\
	&\sim \la e^{i \widehat{\tilde{\phi}}(x,\tau)}e^{ -i \widehat{\tilde{\phi}}(x,0)} \ra 
	=  \exp[ - \la \widehat{\tilde{\phi}}(x,\tau) \widehat{\tilde{\phi}}(x,0) \ra].
\end{align}
Here, the static optimal $\tilde{\theta}_0(x)$ cancels between the two electron operators and the 
small $\delta \tilde{\theta}(x,\tau)$ is neglected compared to $\tilde{\phi}$.  
 Thus, the problem reduces to understanding the correlation function of $\tilde{\phi}$ in the disordered regime.

The approximate effective Hamiltonian is given by 
\be
H_{\rm eff} = \frac{v_F}{4\pi \nu} \int dx ( (\p_x \delta \tilde{\theta})^2 + (\p_x \tilde{\phi})^2) 
+ \int dx  V_0 (\delta \tilde{\theta})^2/2.
\ee
The resulting effective action (in imaginary time) becomes
\be 
\label{dis-action}
S_{\rm eff} =\frac{1}{2\nu} \int d\tau \, 
\int_0^{L_0} dx  \Big [  -\frac{i}{\pi} \p_x \delta \tilde{\theta}  \p_\tau \tilde{\phi} \Big ]  +\int d t H_{\rm eff},
\ee
and the partition function  of $\delta \tilde{\theta}$ and $\tilde{\phi}$ is given by the following path integral:
\be
\label{partition-theta}
Z = \int D[\delta \tilde{\theta},\tilde{\phi}]  \, e^{-S_{\rm eff}}.
\ee
Since the $\tilde{\phi}$ phase plays an important role, 
 we take $\tilde{\phi}$ as the fundamental field
and  integrate out the Gaussian fluctuations of $\delta \tilde{\theta}$ in the path integral calculation, which leads to
an effective action for $\tilde{\phi}$:
\begin{align}
	S_{\rm eff}(\tilde{\phi})&=
	\int d\tau dx dx' 
	(-\frac{1}{\pi})^2 \p_\tau \tilde{\phi}(x,\tau) \Big (  \p_x \p_{x'} \frac{\pi}{4 v_F m  \nu } e^{-m|x-x'|}   \Big )  \p_\tau \tilde{\phi}(x',\tau) 
	\nonumber \\
	&+ \frac{v_F}{4\pi \nu }\int d\tau dx 
	(\p_x  \tilde{\phi})^2,
\end{align}
where 
\be
m = \sqrt{ 2 \nu V_0 \pi/v_F}.
\ee

By decomposing $x$ and $x'$ 
  into the center-of-mass coordinate
  and the relative coordinate, we obtain the following result  for the effective action in frequency-momentum space:
\be 
S_{\rm eff}(\tilde{\phi})=\int \frac{d \omega}{2\pi}  \frac{dq}{2\pi}
(-\frac{1}{\pi})^2 \omega^2 \tilde{\phi}(q,\omega)  \tilde{\phi}(-q,-\omega)   m^2 \frac{\pi}{4 v_F m \nu} 
\Pi(q)+ \frac{v_F}{4\pi \nu}\int d \tau dx 
(\p_x \tilde{\phi})^2,
\ee
where the $\Pi(q)$ is given by
\be
\Pi(q) =
\frac{ m - e^{-L_0 m /2}  m \cos \frac{L_0 q}{2} +e^{-L_0 m /2}  q \sin \frac{L_0 q}{2} }{m^2+q^2}.
\ee
The correlation function of $\tilde{\phi}$ can then be obtained by inverting the kernel of $S_{\rm eff}(\tilde{\phi})$:
\be
\la \tilde{\phi}(x,\tau) \tilde{\phi}(x,0) \ra = \int \frac{dq d\omega}{(2\pi)^2} 
\frac{ e^{-i \omega \tau}}{ \omega^2 m \Pi(q)/2 \pi \nu v_F + v_F q^2/2\pi \nu}.
\ee
In the momentum 
$q$-integral of this expression, the low-
$q$ region is important, allowing us to approximate $\Pi(q)\approx \Pi(0)$ 
and to perform the $q$-integral.
 In the long time limit, we can obtain
\be
\la \tilde{\phi}(x,\tau) \tilde{\phi}(x,0) \ra \approx \int d \omega  K \frac{e^{-i \omega \tau}}{|\omega|} 
\sim  K \ln \tau,
\ee
where $K$ is a certain complicated function of $L_0, m,\nu$ that  parametrizes the approximation done in $q$-integral above.
The right-moving electron Green's function behaves as
\be
G_{e,R}(x,x, \tau)   \sim  \frac{1}{\tau^{K}}.
\ee
Consequently,  in the limit $\omega\rightarrow 0$, the tunneling DOS is given by (with analytic continuation back to real frequency)
\be
D(\omega)  = - \mathrm{Im}  \int d \tau G_{e,R}(x,x, \tau)  \Big \vert_{i \omega \to \omega+ i 0^+} \sim    \omega^{K-1},
\ee
where
\be K \sim \frac{ \pi \nu}{\sqrt{1-e^{-L_0 m}}}. 
\ee
In the limit of a short localization length, $L_0\rightarrow  a_0$,  the exponent $K$ decreases with increasing $V_0$.  This is in qualitative agreement with the numerical results.  For $K>2$, a soft gap is present for $E\approx 0$, in agreement with  Fig.~\ref{fig:dos_and_q1_largeGamma}.  For the critical value $K_c \approx 2$, a linear DOS may emerge, also in agreement with Fig.~\ref{qA1}.  

The derivation presented in this section is not intended for a quantitative comparison with the numerical data. As we mentioned before, it serves as a simple model to enhance our understanding of the role of instantons in the DOS.

\section{Universality Class: Similarities and Differences with a Zigzag Graphene Nanoribbon}\label{section:simanddiff}

What is the relationship between a two-leg ladder and a graphene zigzag nanoribbon~\cite{Ruffieux2016, Kolmer2020, houtsma2021atomically}, which is also a topologically ordered insulator with fractional charges $e/2$~\cite{jeong2019,yang2020,yang2021}?

\begin{figure}[hbt!]
	\begin{center}
		\includegraphics[width = 0.5\textwidth]{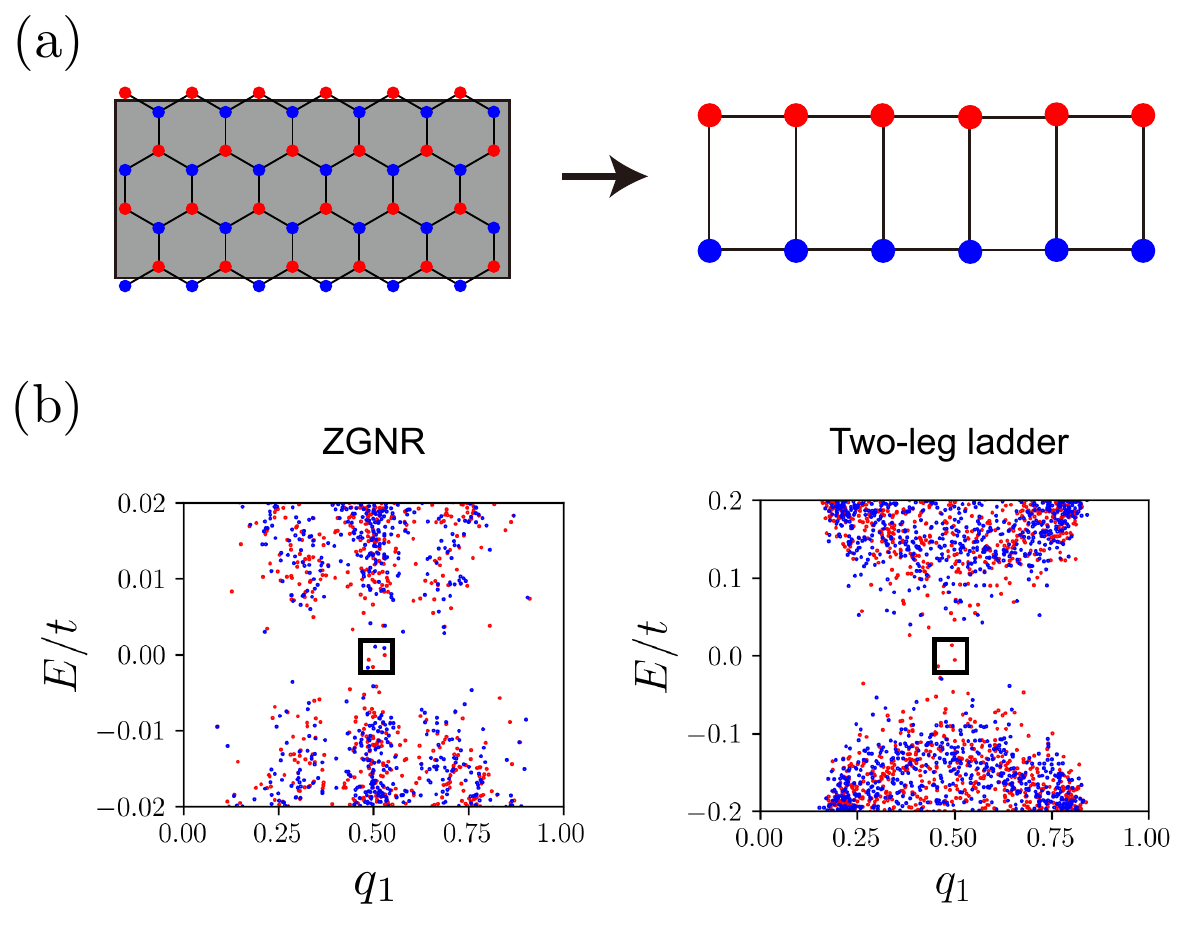}
		\caption{(a) Mapping a zigzag graphene nanoribbon lattice with a gap onto a two-leg ladder is achieved using the Grassmann path integral, where the internal (non-boundary) sites are integrated out. Even if the original lattice has only nearest-neighbor hopping, the mapped two-leg ladder includes longer-range hoppings. {\color{black} (b) \((q_1, E)\) diagrams for a zigzag graphene nanoribbon and a two-leg ladder. For the zigzag graphene nanoribbon, \(q_1\) denotes the charge distribution on the A-sublattice of carbon atoms. The parameters \((U, \Gamma) = (t, 0.15t)\) place the system within the topological order regime, and the diagram is computed by making 700 disorder realizations. For the two-leg ladder, the diagram is computed using \((U, \Gamma, t') = (3t, 0.3t, 0.5t)\) with \(N_D = 6000\) disorder realizations.}}
		\label{Mapping}
	\end{center}	
\end{figure}

The mapping of a graphene zigzag nanoribbon to a two-leg ladder can be achieved by integrating the bulk degrees of freedom {\color{black}[see Fig.~\ref{Mapping}(a)]}. 
This \textit{integrating out} process can be directly implemented in the Grassmann path integral representation of Hartree-Fock  Hamiltonian. (An exhaustive treatment of the path-integral approach for many-particle systems can be found in \cite{Negele}.) We use the mean-field Hamiltonian of the graphene zigzag nanoribbon, given by the standard Hubbard model with on-site repulsion $U$:
\be
\hamop_{\text{HF}} =
\sum_{\vR,\vR'} \sum_{\sigma=\uparrow,\downarrow} \vcop^\dag_{\vR'\sigma}
\Big[ -t_{\vR' \vR} + \delta_{\vR',\vR} \Delta_{\vR \sigma} + V_\vR \Big] \vcop_{\vR\sigma},
\label{hamilHF}
\ee
where $\vcop_{\vR\sigma}$ and $\vcop^\dag_{\vR\sigma}$ represent the electron destruction and creation operators, respectively. Here, $t_{\vR' \vR}$ is the hopping parameter and $\Delta_{\vR \sigma}$ is the mean-field expectation value.
$\vR$ denotes a lattice site and $V_\vR$ represents the disorder potential.  

 The Grassmann path integral derivation is lengthy and nontrivial. The details are provided in Appendix \ref{sec:grass}.  
The mapping results in a bilinear Hamiltonian because the original mean-field Hamiltonian is bilinear,
\be
\label{hamil}
\hamop =
-\sum_{j,j'} \sum_{\sigma=\uparrow,\downarrow} \,  \widehat{\vd}^{+}_{\vj \sigma}
\tilde{t}_{j j'}  \widehat{\vd}_{\vj \sigma},
\ee
where $\tilde{t}_{j j'} $ is the renormalized hopping parameter of the two-leg ladder. When the site label $j=j'$, it represents the effects of the on-site repulsion, in addition to the diagonal disorder.  The same result can also be derived using an alternative approach, namely the reduced density matrix formalism~\cite{Peschel115}.

The crux of the problem is to find $\tilde{t}_{j j'}$.  They depend in a complex way on the eigenstates of the \textit{original} Hartree-Fock Hamiltonian (\ref{hamilHF}) \textcolor{black}{(see Ref.~\cite{yang2021})}. We have numerically evaluated these properties \ and verified that topologically ordered zigzag graphene nanoribbons map onto a two-leg ladder and exhibit fractional charges of $e/2$. 
In the absence of disorder, the resulting two-leg ladder, described by the Lagrangian in Eq.~(\ref{Lag0}), allows the spins at sites in the
$m=1$ chain to be ferromagnetically aligned, as can those in the
$m=2$ chain. However, the
$m=1$ and
$m=2$ chains themselves are antiferromagnetically coupled, similar to the behavior observed in a topologically ordered graphene zigzag nanoribbon.
This contrasts with a two-leg ladder with only nearest-neighbor hopping, where sites within the same or different chains are antiferromagnetically coupled.

However, the resulting ladder system includes longer-range hoppings in addition to nearest-neighbor hoppings. This difference arises because each edge site in a zigzag graphene nanoribbon is connected to only two neighboring sites, whereas each site in a two-leg ladder is connected to three. Moreover, in the absence of disorder, the band structure of a two-leg ladder is nearly semimetallic, while that of a zigzag graphene nanoribbon is insulating.

Nonetheless, two-leg lattices and zigzag graphene nanoribbons belong to the same universality class~\cite{Pachos_2012}, as they share the same topological entanglement entropy $\beta$ \cite{yang2021}, within numerical uncertainty. \textcolor{black}{Additional similarities include the following:  (A) The exponent in the density of states, $D(E) \propto \left[ e^{\alpha E^{\eta}} - 1 \right]$, is found to be approximately $\eta \approx 2$ in disordered two-leg ladders and graphene nanoribbons \{see Fig. 4 of the current manuscript and Eq.~(8) of Ref.~\cite{yang2020}\}. (B) The $(q_1, E)$ plots in Fig.~\ref{Mapping}(b) for disordered two-leg ladders and graphene nanoribbons exhibit  strikingly similar features near the critical point $(q_1, E)=(1/2,0)$, where accurate fractionalization is expected to emerge in the thermodynamic limit.}
We propose that two-leg electron ladders could serve as a model for studying a topologically ordered system with fractional charges of $e/2$.

\section{Discussion and Conclusions}\label{section:discussion}

We numerically studied the Hubbard model of disordered two-leg electron ladders. Our findings reveal that when an exponentially decaying gap is well developed in the DOS, magnetic domain walls form, leading to fractional charges residing on these walls. In this regime, our mean-field approach demonstrates that the two-leg electron ladders exhibit a well-defined, universal topological entanglement entropy. The validity of the mean-field approach is supported by the suppression of quantum fluctuations because of the gap, as demonstrated in a previous study based on the matrix product representation of density matrix renormalization group studies. When a soft gap is not well developed, our results suggest that the system resides in a crossover regime where quasi topological order emerges, with the topological entanglement entropy exhibiting significant variance.

To better understand the role of magnetic order, kinks, and instantons in the numerical results, we discussed the Shankar-Witten-type Lagrangians.  Then, inspired by the Lagrangian that describes the edge of a fractional quantum Hall state \cite{Wen1992}, we proposed an effective bosonic Lagrangian to describe semion excitations. In this model, the instanton operator is mathematically well-defined when two semions reside on opposite chains, meaning their fusion produces a fermion (instanton). Using the bosonization method, we then computed the shape of the electronic DOS within the framework of the pinned-CDW model. For weaker disorder, our semion model produces a soft gap, while at a critical disorder strength, it results in a linear density of states, qualitatively agreeing with our numerical results.

We demonstrated that a zigzag nanoribbon cannot be mapped onto a two-leg electron ladder without introducing long-range hopping. This difference arises from the connectivity of their lattice sites: In a two-leg electron ladder, each site is connected to three neighboring sites, whereas each edge site in a zigzag graphene nanoribbon is connected to only two. Moreover, in the absence of disorder, the band structure of two-leg electron ladders is nearly semimetallic, whereas that of zigzag graphene nanoribbons is insulating.
Another key difference is that in zigzag ribbons, disorder acts as a singular perturbation, causing the topological entanglement to change discontinuously from zero to a finite universal value~\cite{yang2021}, unlike in two-leg electron ladders. Despite these differences, both systems exhibit the same fractional charges and nearly identical universal values of topological entanglement entropy, $\beta$, within the numerical uncertainty. Therefore, we believe that two-leg electron ladders and zigzag graphene nanoribbons belong to the same universality class.

Several effects are crucial for understanding the robustness of fractional charges. Electron localization plays a significant role in stabilizing fractional charges.  Upon weak doping,  midgap localized fractional states with 
$E\approx 0$ exist~\cite{yang2022}, but they do \textit{not} spatially overlap with each other—a hallmark of localization~\cite{Altshuler, GV2000}. Furthermore, fractional states are isolated from other states of higher energy by an exponential gap in the DOS. This implies that quantum fluctuations, which are neglected in the Hartree-Fock approximation, have a negligible effect on fractional charges, as indicated by the density matrix renormalization group approach~\cite{yang2022}.
As the strength of disorder and/or interchain coupling deviates from the universal region in the parameter space, the variance of the topological entanglement entropy is expected to grow, leading to crossover phases of quasitopological order~\cite{Le_2024}.  In this case, fractional charges are not well-defined.

It would be interesting to experimentally probe fractional charges. \textcolor{black}{Atomically precise zigzag graphene nanoribbons can now be fabricated~\cite{Ruffieux2016, Kolmer2020, houtsma2021atomically}.  The definitive experimental probe of fractional charges is the shot noise measurement, a well-established method for confirming fractional charge in quantum Hall systems. We propose that the tunneling of fractional charges between the opposite zigzag edges \cite{Le_2024} could be detected via shot noise measurements. These tunneling events are also expected to exhibit a zero-bias anomaly. It may also be worthwhile to investigate the effect of doping. Doping will alter the ground state in a singular manner, transforming it from an antiferromagnetic state to a spin density wave state. The peak value of this anomaly shows an unusual nonlinear dependence on doping, which could serve as a distinctive experimental signature \cite{yang2022}.}

  The ground state of a gas of particles with rational fractional statistics (anyons) is a superfluid, where bosonic degrees of freedom emerge as composites of an integral number of anyons~\cite{Laughlin1988, Canright1989, Zee1990}. Investigating the effects of disorder on anyon superconductivity could be a compelling direction for future research.

\textcolor{black}{The data and programming code of this article are available from Zenodo \cite{zenodo}}.

\begin{acknowledgments}
This work was supported by the Korea Institute of Science and Technology Information (KISTI) Supercomputing Center with supercomputing resources including technical support (KSC-2024-CRE-0400). H.-A.L acknowledges the financial support from the Institute for Basic Science, Republic of Korea (IBS-R027-D1).
\end{acknowledgments}

\bibliography{aapmsamp.bib}

\appendix

\section{Mapping Into Two-Leg Ladder}
\subsection{ Introduction to Grassmann path integral approach}
\label{sec:grass}

A density matrix $e^{-\beta \hat{\mathsf{H}}}$ can be written as a product of factors (the so-called Trotter product formula) over infinitesimal (imaginary)
time slices:
\begin{equation}
	e^{-\beta \hat{\mathsf{H}}} = \lim_{N \to \infty} \, \prod_{N \, \text{times}} e^{-\frac{\beta}{N} \hat{\mathsf{H}}}
	\cdots e^{-\frac{\beta}{N} \hat{\mathsf{H}}}.
	\label{Trotter}
\end{equation}
The partition function is given by $Z = \mathrm{Tr} e^{-\beta \hat{\mathsf{H}}}$.
The path integral representation of the partition function can be obtained by inserting the closure relations of \textit{coherent states} between the factors in Eq.(\ref{Trotter}) (here, the Hamiltonian is assumed to be normal ordered). The coherent state is the eigenstate of the annihilation operator,
\begin{equation}
	\hat{\vc}_\alpha \vert \xi \rangle = \xi_\alpha \vert \xi \rangle. 
\end{equation}
If the operator $\hat{\vc}_\alpha$ is a fermion operator, then the anticommutation relation of 
fermions dictates that the eigenvalues $\xi_\alpha$ must be \textit{anticommuting} Grassmann numbers,
\begin{equation}
	\xi_\alpha \xi_{\alpha'} = - \xi_{\alpha'} \xi_\alpha. 
\end{equation}
The closure relation of fermionic coherent states takes the following form:
\begin{equation}
	\label{closure}
	\mathrm{Identity} = \int \prod_\alpha \, d \xi_\alpha^* \xi_\alpha \, e^{-\sum_\alpha \xi^*_\alpha \xi_\alpha}
	\, \vert \xi \rangle \langle \xi \vert,
\end{equation}
and the matrix element of \textit{normal ordered} operator $\widehat{\mathsf{A}}$ is given by
\begin{equation}
	\label{matrixelement}
	\langle \xi \vert \widehat{\mathsf{A}}(\vc^\dag_\alpha, \vc_\alpha) \vert \xi' \rangle  = e^{\sum_\alpha \xi^*_\alpha 
		\xi_\alpha'}\, \mathsf{A}(\xi^*_\alpha,\xi'_\alpha).
\end{equation}
Using Eqs.~(\ref{closure}) and (\ref{matrixelement}), 
the partition function of a fermionic Hamiltonian $\widehat{\mathsf{H}}(\hat{\vc}^\dag, \hat{\vc})$ can be expressed 
in the form of a
Grassmann path integral,
\begin{equation}
\label{gPath}
	Z = \int D [\xi^*_\alpha(\tau),\xi_\alpha(\tau)]\,
	\exp \Big[-\int_0^\beta d \tau \, \big( \sum_\alpha \xi^*_\alpha(\tau)(\frac{\partial}{\partial \tau}-\mu )
	\xi_\alpha(\tau) +  \mathsf{H}(\xi^*_\alpha,\xi_\alpha) \big )\Big ],
\end{equation}
where the Grassmann variables $\xi^*_\alpha(\tau),\xi_\alpha(\tau)$ satisfy the antiperiodic temporal boundary condition
$\xi_\alpha(\beta) =-\xi_\alpha(0)$, and $\mu$ is the chemical potential.

\subsection{Green's function of two-leg ladder}

We can map the original system onto a subsystem by integrating out the variables $\xi_{\alpha}(\tau)$
 with indices $\alpha$
that are exterior to the subsystem. The partition function of the fermion system is then given by the following Grassmann path integral from Eq.~(\ref{gPath}):
\be
\label{partition0}
Z = \int D[\bar{\vc}_{\vR \sigma}(\tau),\vc_{\vR \sigma}(\tau)] \, e^{-S[\bar{\vc},\vc]/\hbar},
\ee
where the action in imaginary time is given by 
\be
\label{partition1}
S[\bar{\vc},\vc] = \int_0^{\hbar \beta} \, d \tau \, \sum_{\vR,\sigma}  \bar{\vc}_{\vR \sigma}(\tau) \hbar \frac{\p  \vc_{\vR \sigma}(\tau)}{\p \tau}
+ \int_0^{\hbar \beta} d \tau \,  \ham(\bar{\vc}_{\vR \sigma}(\tau),\vc_{\vR \sigma}(\tau) ).
\ee
Note that the action $S[\bar{\vc},\vc]$ is \textit{quadratic} in the Grassmann variables
$\bar{\vc}_{\vR \sigma}(\tau),\vc_{\vR \sigma}(\tau)$
for the case of  HF Hamiltonian~(\ref{hamil}), so that the partition function can be evaluated exactly using Grassmann Gaussian integration.
From now on, we set $\hbar =1$. 
In our approach, the following Grassman representation of the Dirac delta function plays a pivotal role:
\be
\label{delta}
\delta(\xi,\xi') = \int d \lambda \,  e^{-\lambda (\xi-\xi')} = -(\xi - \xi'),
\ee
where $\xi,\xi'$ and $\lambda$ are Grassmann variables, and
$\lambda$ plays the  role of a \textit{Lagrange multiplier}.

Let $\mathsf{J}$ be the index set for the subsystem sites (for example, the two chains of a two-leg ladder). Then, the following integral over the auxiliary Grassmann variables of the subsystem
$\bar{\vd}_{\mathsf{j}}, \vd_{\mathsf{j}}$ is unity (spin indices suppressed for clarity):
\be
\label{step1}
\int \, \prod_{\mathsf{j} \in \mathsf{J} }  D[ \bar{\vd}_{\mathsf{j}}, \vd_{\mathsf{j}}] \,
\delta(\bar{\vd}_{\mathsf{j}},\bar{\vc}_{\mathsf{j}}) \,  \delta(\vd_{\mathsf{j}},\vc_{\mathsf{j}})  = 1.
\ee
The above delta function integrals can be exponentiated using Eq.~(\ref{delta}), with a generalization to multiple sites, leading to the following:
\begin{align}
	\label{step3}
	Z &= \int D[\bar{\vc}_{ \vR \sigma}(\tau),\vc_{ \vR \sigma }(\tau)] \,
	\prod_{\mathsf{j} \in \mathsf{J} }  D[ \bar{\vd}_{\mathsf{j} \sigma}(\tau), \vd_{ \mathsf{j} \sigma }(\tau)]
	D[ \bar{\lambda}_{ \mathsf{j} \sigma }(\tau), \lambda_{ \mathsf{j}  \sigma}(\tau)] \nonumber \\
	& \times 
	\exp \Big [- S -\sum_\tau \,\sum_{\mathsf{j},\sigma} \Big\{-( \bar{\vd}_{ \mathsf{j} \sigma}(\tau)-\bar{\vc}_{ \mathsf{j} \sigma}(\tau) )\lambda_{ \mathsf{j} \sigma }(\tau) +  \bar{\lambda}_{ \mathsf{j} \sigma }(\tau) ( \vd_{ \mathsf{j} \sigma}(\tau)-\vc_{ \mathsf{j} \sigma}(\tau) ) 
	\Big \} \Big ],
\end{align} 
The Grassmann integral over $ \bar{\vc}_{\vR \sigma}(\tau),\vc_{ \vR \sigma }(\tau) $
can be performed exactly using the Grassmann Gaussian integral formula 
\{see Eq. (1.184) of Ref.~\cite{Negele}\}.
It turns out that the integral over the Grassmann Lagrange multipliers 
$\bar{\lambda}_{ \mathsf{j} \sigma }(\tau), \lambda_{ \mathsf{j} \sigma}(\tau)$ is also Gaussian, so  they can 
be also integrated out exactly, allowing us to obtain an effective action  for the subsystem described by 
$\bar{\vd}_{\sigma \mathsf{j} }(\tau), \vd_{\sigma \mathsf{j} }(\tau)$ explicitly.

It turns out that it is necessary to first find the Green's function 
$G_\sigma$ of the original system:
\be
\label{greenZGNR}
\sum_{ \vR^{\prime \prime}}
\Big[  \delta_{\vR' \vR^{\prime \prime}} \frac{\p}{\p \tau'} -t_{\vR' \vR^{\prime \prime}} +\delta_{\vR',\vR^{\prime \prime}} \Delta_{\vR^{\prime \prime} \sigma} \Big ][-G_\sigma]_{\vR^{\prime \prime} \tau^{\prime \prime}; \vR, \tau} =
\delta(\tau'-\tau) \delta_{\vR',\vR},
\ee
where $\tau,\tau'$ are the \textit{imaginary time} variables employed in the path integral formulation of the partition function.

Following this approach,
the partition function for a subsystem  \{indexed by $\mathsf{j} \in \mathsf{J}$
and with $\bar{\vd}_{\mathsf{j} \sigma}(\tau), \vd_{\mathsf{j} \sigma}(\tau)$ being the corresponding Grassmann variables\}, described by the HF Hamiltonian Eq.~(\ref{hamil}), can be obtained as follows:
\begin{align}
	\label{step11}
	Z_{\rm sub} & =\int \prod_{\mathsf{j} \in \mathsf{J} }  D[ \bar{\vd}_{\mathsf{j} \sigma}(\tau), \vd_{\mathsf{j} \sigma}(\tau)]  \nonumber \\
	&\times  \det_{\mathsf{J}} [-G_\sigma ]\, 
	\exp \Big \{ +\sum_\sigma \sum_{\tau',\tau} \sum_{\vj',\vj} 
	(\bar{\vd}_{ \vj' \sigma \tau'} )  [G_\sigma]^{-1,(\mathsf{J})}_{\vj' \tau',  \vj \tau} (\vd_{ \vj \sigma\tau}) \Big \}
\end{align}
It is very important to note that the inverse matrix 
$[G_\sigma]^{-1,(\mathsf{J})}$ of the Green's function of 
Eq.~(\ref{greenZGNR}), \textcolor{black}{used in
 Eq.~(\ref{step11}),} is defined with respect to the subsystem sites only (not over 
the whole original sites $\vR$).
This has been made explicit by the superscript $(\mathsf{J})$.
Therefore, the effective action (in energy-position space) of the subsystem is given by
\be
\label{result2}
S^{(\mathsf{d})}_{\mathrm{eff}} =
-\sum_\sigma \sum_{\epsilon} \sum_{\vj',\vj} 
\bar{\vd}_{ \vj' \sigma } (\epsilon) 
[G_\sigma]^{-1 (\mathsf{J}) }_{\vj',  \vj}(i \epsilon) \vd_{ \vj \sigma }(\epsilon).
\ee
Note that the kernel
$[G_\sigma]^{-1 (\mathsf{J}) }_{\vj',  \vj}(i \epsilon)$ includes the 
renormalization effects arising   from  integrating out bulk sites.

In the energy-position basis, the Green's function can be written as 
\be
\label{def-green}
[G_\sigma]_{\vj^{\prime \prime}  \tau^{\prime \prime}, \vj \tau} 
= \frac{1}{\beta}\, \sum_{\epsilon} G_\sigma(i \epsilon)_{\vj^{\prime \prime},\vj} \, e^{-i \epsilon (\tau^{\prime \prime} - \tau)}
=-\la T_\tau \, \widehat{\vc}_{ \vj^{\prime \prime} \sigma}(\tau^{\prime \prime} )
\widehat{\vc}^\dag_{ \vj \sigma}(\tau) \ra.
\ee
The spectral decomposition of $G_\sigma(i \epsilon)_{\vj^{\prime \prime},\vj}$ takes the following form,
\be
\label{green.res}
G_\sigma(i \epsilon)_{\vj^{\prime \prime},\vj} =  \sum_{P}
\frac{\psi_{P \sigma}(\vj^{\prime \prime })   \psi_{P \sigma}^*(\vj) }{i\epsilon-E_P},
\ee
where $P$ labels the eigenstates of the original system, and $\psi_{P\sigma}(\vj)$ is the lattice eigenfunction
of the Hamiltonian of the original system.

Taking the imaginary time coordinates of Eq.(\ref{def-green}) as 
$ \tau = \tau^{\prime \prime}+0^+$, we can obtain the 
\textit{equal time Green's function},
\be
\label{correlation}
	C^\sigma_{\vj \vj^{\prime }} \equiv \la  \widehat{\vc}^\dag_{\vj \sigma}(\tau)   \widehat{\vc}_{ \vj^{\prime } \sigma}(\tau)  \ra =
	\la  \bar{\vd}_{\vj \sigma}(\tau)  \vd_{ \vj^{\prime } \sigma}(\tau)  \ra = 
	\sum_{P}\, \psi_{P\sigma}(\vj^{\prime  })   \psi_{P\sigma}^*(\vj) \, n_F(E_{P_\sigma}),
\ee
where $n_F$ is the Fermi occupation number, which is obtained by summing over the Matsubara frequency  $i \epsilon$ in Eq.~(\ref{green.res}).
Note that the first bracket of Eq.~(\ref{correlation}) denotes  an operator average, 
while the second bracket is to be understood in the context of path integral.
We emphasize again that the correlation function Eq.~(\ref{correlation}) has been obtained 
in terms of the eigenfunctions and the eigenvalues of the \textit{original} system.

\subsection{Effective  Hamiltonian of two-leg ladder}

Next, consider the following effective Hamiltonian \textit{defined on the subsystem} ($\widehat{\vd}^\dag_{ \vj' \sigma}$, $\widehat{\vd}_{\vj \sigma}$ are the corresponding creation/annihilation operators, and $\vert \mathsf{J} \vert$ denotes the number of subsystem sites):
\be
\label{eff1}
\hamop^\sigma_{\rm eff} = \sum_{\vj',\vj}^{\vert \mathsf{J} \vert} \,\widehat{\vd}^\dag_{ \vj' \sigma}
\ham^\sigma_{\vj' \vj}  \widehat{\vd}_{\vj \sigma},
\ee
where $\ham^\sigma_{\vj' \vj'}$ is a $\vert \mathsf{J} \vert \times \vert \mathsf{J} \vert $
(numerical) Hermitian matrix 
that is, at this stage, unknown.
The goal is to obtain (or express) the correlation function in Eq.~(\ref{correlation}) in terms of the matrix elements $\ham^\sigma_{\vj' \vj}$.
Remembering that the correlation function in Eq.~(\ref{correlation}) has been obtained explicitly in terms of the eigenfunctions and eigenvalues of the original system, we should be able to find an explicit relation between the parameters of the original system and those of the reduced system consisting of the chains that form the two-leg ladder.

The partition function corresponding to the effective Hamiltonian in Eq.~(\ref{eff1}) is given by the following Grassmann path integral:
\begin{align}
	\label{eff2}
	Z_{\rm eff} & =  \mathrm{Tr} e^{-\beta \sum_\sigma \hamop^\sigma_{\rm eff}  } \nonumber \\
	&=
	\int \prod_{\vj \in \mathsf{J}} D[\bar{\vd}_{\vj \sigma}(\tau),\vd_{ \vj \sigma}(\tau)] e^{-S_{\rm eff}[\bar{\vd},\vd] }.
\end{align}
In the same way as in Eqs.~(\ref{partition0}) and (\ref{partition1}), we can obtain the effective action 
$S_{\rm eff}[\bar{\vd},\vd]$,
\begin{align}
	\label{eff3}
	S_{\rm eff}[\bar{\vd},\vd] &= \int_0^{\hbar \beta} \, d \tau \, \sum_{\vj,\sigma}  \bar{\vd}_{\vj \sigma}(\tau) \hbar \frac{\p  \vd_{\vj \sigma}(\tau)}{\p \tau}
	+ \int_0^{\hbar \beta} d \tau \,  \ham(\bar{\vd}_{\vj \sigma}(\tau),\vd_{\vj \sigma}(\tau) )
	\nonumber \\
	&=\int_0^{\hbar \beta} \, d \tau \, \sum_{\vj,\sigma}  \bar{\vd}_{\vj \sigma}(\tau) \hbar \frac{\p  \vd_{\vj \sigma}(\tau)}{\p \tau}
	+ \int_0^{\hbar \beta} d \tau \,\sum_\sigma 
	\sum_{\vj',\vj}^{\vert \mathsf{J} \vert} \,\vd^\dag_{ \vj' \sigma}
	\ham^\sigma_{\vj' \vj}  \vd_{ \vj \sigma}.
\end{align}

Consider the eigenvalues and eigenfunctions of the Hermitian matrix
$\ham^\sigma_{\vj' \vj}$,
\be
\label{eff-eigen}
\sum_{ \vj}  \ham^{\sigma}_{\vj' \vj}  \Phi^{q \sigma}_{\vj} 
=\epsilon_{q \sigma}   \Phi^{q \sigma}_{\vj'},
\ee
where $\epsilon_{q \sigma} $ is the eigenvalue labeled by the index $q\sigma$
(which ranges up to $\vert \mathsf{J} \vert$), and 
$  \Phi^{q \sigma}_{\vj'}$ is the corresponding real-space eigenfunction. It is important to note that this is a formal procedure, as the matrix 
 $\ham^\sigma_{\vj' \vj}$ is not yet known.
We emphasize that these eigenvalues and eigenfunctions are \textit{not}
identical with those of the original system appearing in Eq.(\ref{correlation}).
Employing the Gaussian Grassmann integral formula \{Eq. (1.184) of \cite{Negele}\},
we obtain the correlation function
$-\la \vd_{\vj \sigma}(\epsilon)  \bar{\vd}_{\vj' \sigma}(\epsilon) \ra$,
\be
\label{eff4}
-\la \vd_{\sigma \vj}(\epsilon)  \bar{\vd}_{\sigma \vj'}(\epsilon) \ra = 
\Big[ i \epsilon \, \mathrm{I}- \ham^\sigma \Big ]^{-1}_{\vj,\vj'}
\equiv C^\sigma_{\vj' \vj}(i \epsilon),
\ee
where \( \mathrm{I} \) is the identity matrix in the $\vj,\vj' $ indices.
Clearly, the two matrices $ \ham^\sigma$ and $C^\sigma$ can be diagonalized simultaneously.
The spectral decomposition of the Hamiltonian matrix achieves this diagonalization,
\be
\label{spectral-decomposition}
\la \vj \vert  \ham^\sigma \vert \vj' \ra = \sum_{q\sigma}
\,\epsilon_{q \sigma} \Phi^{q \sigma}_\vj  (\Phi^{q \sigma}_{\vj'})^*
\ee
Applying Eq. (\ref{spectral-decomposition}) to the correlation function in Eq. (\ref{eff4}),
we obtain
\be
\label{eff5}
C^\sigma_{\vj' \vj}(i \epsilon) = \la  \bar{\vd}_{\vj' \sigma}(\epsilon) \vd_{\vj \sigma}(\epsilon)  \ra
=\sum_{q \sigma} \frac{  (\Phi^{q \sigma}_{\vj'})^* \Phi^{q\sigma}_{\vj} }{ i \epsilon -\epsilon_{q \sigma}}.
\ee
The frequency summation in Eq. (\ref{eff5})
(with the time ordering factor $e^{i 0^+}$ 
  implicitly understood)
yields the \textit{equal-time correlation function},
which should be identical to that in Eq. (\ref{correlation}):
\be
\label{result3}
	\frac{1}{\beta}\,\sum_{i \epsilon} \,C^\sigma_{\vj' \vj}(i \epsilon) = \la  \bar{\vd}_{\sigma \vj'}(\tau) \vd_{\sigma \vj}(\tau)  \ra
	=\sum_{q \sigma}^{\vert \mathsf{J} \vert}  (\Phi^{q \sigma}_{\vj'})^* \Phi^{q\sigma}_{\vj} \, n_F(\epsilon_{q\sigma}) =C^\sigma_{\vj' \vj}=
	\text{Eq.(\ref{correlation})}.
\ee
Equation (\ref{result3}) establishes a direct mapping between the original system and the reduced subsystem.

Next, using the Fermi-Dirac distribution $n_F(\epsilon_{q\sigma}) =  [1+ e^{\beta \epsilon_{q\sigma}}]^{-1}$, we obtain
\be
\label{occupation}
[n_F(\epsilon_{q\sigma})]^{-1} -1 =  e^{ \beta \epsilon_{q\sigma}}  
\to \epsilon_{q\sigma} = \frac{1}{\beta}\, \ln \frac{1-n_F(\epsilon_{q\sigma})}{n_F(\epsilon_{q \sigma})}.
\ee
Substituting Eq.~(\ref{occupation}) into the spectral decomposition Eq.~(\ref{spectral-decomposition}), we obtain
\be
\label{hamil-corr}
\la \vj \vert  \ham^\sigma \vert \vj' \ra = \sum_{q\sigma}
\Phi^{q \sigma}_\vj  (\Phi^{q \sigma}_{\vj'})^*
\frac{1}{\beta}\, \ln \frac{1-n_F(\epsilon_{q\sigma})}{n_F(\epsilon_{q \sigma})}
\ee
The above result is spectrally decomposed, allowing it to be rewritten in terms of the equal-time correlation function using Eq.~(\ref{result3}):
\be
\label{result4}
	\la \vj \vert  \ham^\sigma \vert \vj' \ra
	=\frac{1}{\beta}\, \left [  \ln \frac{ \mathrm{I}-C^\sigma}{C^\sigma} \right ]_{\vj',\vj} 
	\to [\ham^\sigma ]^T = \frac{1}{\beta}\, \left [  \ln \frac{ \mathrm{I}-C^\sigma}{C^\sigma} \right ]
	.
\ee
This result is identical to Eq.~(12) of Ref.~\cite{Peschel115}.  Recall that the equal-time correlation function has been explicitly computed in terms of the original Hamiltonian, as shown in 
Eq.~(\ref{correlation}). Therefore, we have obtained an explicit mapping between the parameters of the original system and those of the reduced subsystem.

\label{sec:grass-reduced}


\end{document}